\documentclass[reprint,prl,aps,amsmath,amssymb,superscriptaddress,nofootinbib]{revtex4-2}
\usepackage{graphicx}
\usepackage[english]{babel}
\usepackage[colorlinks, citecolor=blue, urlcolor=blue, linkcolor=red,breaklinks]{hyperref}
\usepackage{physics}
\usepackage{xcolor}
\usepackage[normalem]{ulem}
\usepackage{float}
\usepackage{tabularx}
\usepackage{xr}

\usepackage{multirow}
\usepackage{CJK}
\begin{document}

\title{Evolution of Superconductivity in Twisted Graphene Multilayers}

\author{Min Long}
\thanks{These authors contributed equally}
\affiliation{IMDEA Nanoscience, Faraday 9, 28049 Madrid, Spain}
\affiliation{Department of Physics and HKU-UCAS Joint Institute of Theoretical and Computational Physics, The University of Hong Kong, Pokfulam Road, Hong Kong SAR, China}

\author{Alejandro Jimeno-Pozo}
\thanks{These authors contributed equally}
\affiliation{IMDEA Nanoscience, Faraday 9, 28049 Madrid, Spain}

\author{H\'ector Sainz-Cruz}
\affiliation{IMDEA Nanoscience, Faraday 9, 28049 Madrid, Spain}

\author{Pierre A. Pantale\'on}
\email{pierre.pantaleon@imdea.org}
\affiliation{IMDEA Nanoscience, Faraday 9, 28049 Madrid, Spain}

\author{Francisco Guinea}
\affiliation{IMDEA Nanoscience, Faraday 9, 28049 Madrid, Spain}
\affiliation{Donostia International Physics Center, Paseo Manuel de Lardiz\'abal 4, 20018 San Sebastián, Spain}

\date{\today}

\begin{abstract}
The group of moiré graphene superconductors keeps growing, and by now it contains twisted graphene multilayers as well as untwisted stacks. We analyze here the contribution of long range charge fluctuations in the superconductivity of twisted double graphene bilayers, helical trilayers, and other multilayers, and compare the results to twisted bilayer graphene. A diagrammatic approach which depends on a few, well known parameters is used. We find that the critical temperature and the order parameter differ significantly between twisted double bilayers and helical trilayers on one hand, and twisted bilayer graphene on the other. This trend, consistent with experiments, can be associated to the role played by moiré Umklapp processes in the different systems.

\end{abstract}

\maketitle

{\it Introduction.} The discovery of unconventional superconductivity (SC) in twisted graphene stacks, including twisted bilayer graphene (TBG)~\cite{Cao2018,Yankowitz2019,Lu2019_SC_TBG,Stepanov2020_TBG,Oh2021evidence}, twisted trilayer graphene (TTG)~\cite{Park2021_SC_TTG, Hao2021_SC_TTG,Kim2022evidence,Liu2022Isospin} and twisted multilayers \cite{Park2022_multi, Zhang2022_promotionSC}, as well as their non-twisted counterparts such as Bernal bilayer graphene (BBG)~\cite{zhou2022isospin,zhang2022spin,holleis2023SC} and rhombohedral trilayer graphene (RTG)~\cite{Zhou2021SuperRTG}, has sparked considerable interest. The rich phase diagrams of these materials, together with the possibility to switch between phases by changing the carrier density in-situ has been celebrated as the beginning of a new era in materials science~\cite{Andrei2020Graphene,Balents2020Superconductivity}. The ground-breaking finding of superconductivity in TBG reignited the quest for superconductivity in other graphene systems. The most clear candidate was twisted double bilayer graphene (TDBG), consisting of two Bernal bilayers with a relative twist. However, superconductivity in TDBG remained elusive despite extensive searches that revealed numerous correlated phases and symmetry-broken states~\cite{Burg2019Correlated,Shen2020Correlated,Cao2020SpinPolarized,Liu2020SpinCorrelated,He2020SymBreak,Kuiri2022Spontaneous, Hsu2020TopoSC}. Remarkably, a recent experiment has shown that TDBG is indeed a superconductor~\cite{Su2023Superconductivity}, with a critical temperature of $T_c\approx 65$ mK. Weak superconductivity has also been observed in alternating-twist trilayer graphene with unequal angles~\cite{Uri2023Superconductivity}. A rich phase diagram, but no superconductivity, has been found in helical TTG (hTTG), for which the twist angles are equal in both magnitude and direction~\cite{Nakatsuji2023Multi,Devakul2023Magic,Guerci2023Chern,Mao2023Supermoire,Zhu2020Twisted,Mora2019Flatbands,Popov2023Stronly}. TDBG shares some features with other graphene superconductors, such as the emergence of superconductivity near van-Hove singularities and in the vicinity of flavor symmetry-broken phases. However, the critical temperature for TBG and the alternating-twist family is in the range 1-2 K~\cite{Cao2018,Yankowitz2019,Lu2019_SC_TBG,Stepanov2020_TBG,Oh2021evidence,Park2021_SC_TTG,Hao2021_SC_TTG,Kim2022evidence,Liu2022Isospin,Park2022_multi,Zhang2022_promotionSC}, whereas TDBG exhibits a lower critical temperature of $T_c\approx 65$ mK~\cite{Su2023Superconductivity}, comparable to RTG and BBG~\cite{Zhou2021SuperRTG,zhou2022isospin,zhang2022spin,holleis2023SC,Pantaleon2023ReviewSC}.

Many theories have been proposed to describe SC in moiré graphene, but the subject remains a topic of debate. It has been argued that direct electron-phonon interactions could account for superconductivity~\cite{Peltonen2018, Wu2018, Choi2018, Lian2019, Wu2019, Schrodi2020,Wu2020FerroSC, Li2020PhononSC, Samajdar2020SC,Choi2021,Qin2023,chou2024topological}. There is significant evidence that points to an unconventional mechanism and pairing symmetry, such as
resilience to magnetic fields well above the limit for paramagnetic superconductors~\cite{cao2021pauli}, zero-bias subgap conductance peaks due to Andreev reflection~\cite{Oh2021evidence, Kim2022evidence} and non-reciprocal Josephson currents~\cite{lin2022zero}, which are proof of broken time-reversal symmetry in the superconducting phase. These observations, together with the fact that SC emerges only in narrow bands which maximize electronic interactions, have triggered enormous interest in unconventional, purely electronic mechanisms~\cite{Gonzlez2019, Roy2019, Goodwin2019, Lewandowski2021, Sharma2020, Samajdar2020SC, cea21Coulomb,Pahlevanzadeh2021DMFT, Crepel2022Unconventional, Cea2023Superconductivity,Gonzlez2023TTG}. Other scenarios focus on quantum fluctuations~\cite{Po2018, You2019,Lee2019SpinTriplet, Wu2020, Kumar2021, Kozii2022,Fischer2022TTG}, or van Hove singularities at the Fermi level~\cite{Isobe2018, Sherkunov2018, Chichinadze2020, Lin2020}.

\begin{figure}[t]
    \centering
    \includegraphics[width=8.5cm]{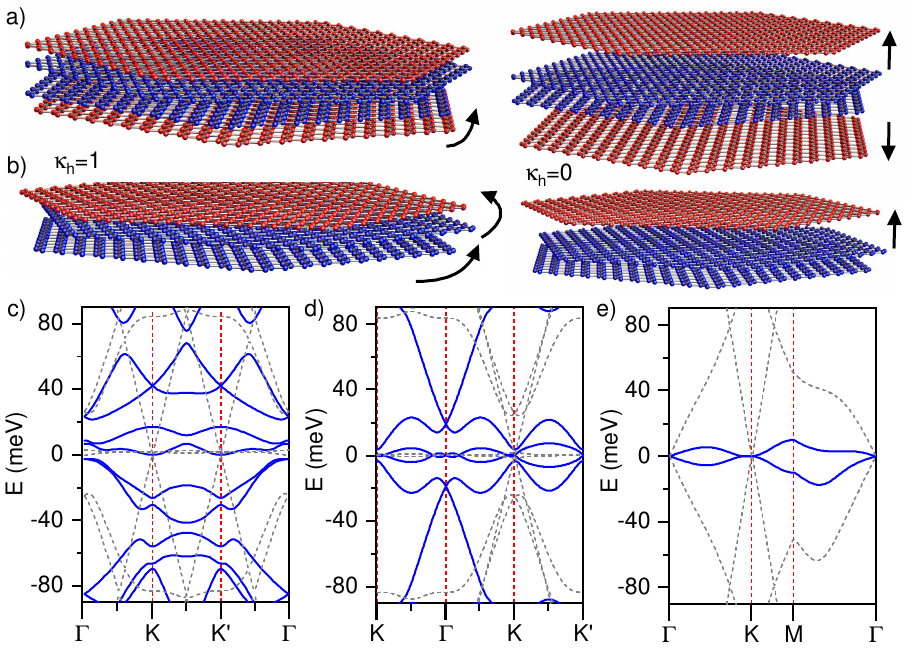}
    \caption{Schematic view of the real space configuration for a transition from a) TDBG ($\kappa_{h}=1$) to TBG + 2 MG ($\kappa_{h}=0$) and from b) hTTG ($\kappa_{h}=1$) to TBG + 1 MG ($\kappa_{h}=0$). The corresponding band structures are show in c) for TDBG ($\theta=1.10^{\circ}$), d) hTTG ($\theta=1.10^{\circ}$) and e) hTTG ($\theta=1.90^{\circ}$). Color scheme is such that blue and gray lines are for $\kappa_{h}=1$ and $0$, respectively. In a) and b) we set $\theta=10^{\circ}$ for illustrative purposes. Curved arrows indicate the twist angle directions, and vertical arrows indicate the decoupling direction. The layers are distinguished by color only for visualization purposes, and red layers are those being separated by the coupling parameter $\kappa_{h}$.}
    \label{fig:1}
\end{figure}

Here we investigate superconductivity in TBG, TDBG and hTTG mediated by the screened long-range Coulomb interaction. We use a diagrammatic approach based on the seminal Kohn-Luttinger (KL) analysis~\cite{Kohn1965, Chubukov1993KL2D, cea21Coulomb}, which includes direct electronic interactions as well as phonon-mediated electronic interactions, and takes into account moiré induced Umklapp processes. We apply this framework to the continuum model of both TDBG and hTTG, including a coupling scaling parameter, $\kappa_{h}$, that allows us to continuously convert them to a TBG plus decoupled graphene monolayers (MG), as sketched in Fig.~\ref{fig:1}. For TBG, we find a critical temperature of $T_{c}\approx 2$~K. We then track the evolution of superconductivity along the transition towards TDBG and observe that superconductivity is suppressed before reaching the TDBG limit. Upon introducing an electric field, we observe that while the critical temperature decreases in TBG, in TDBG it appears and reaches a maximum of $T_{c}\approx 10$ mK. For hTTG, we predict superconductivity with a maximum $T_{c}\approx 80$ mK in absence of any electric field. We find that Umklapp phonon dressing is what makes TBG stand out with respect to the other stacks. 

The paper is organized as follows: first we describe the KL-RPA mechanism and then analyze the superconducing properties of TBG, TDBG and hTTG at magic-angle. Then we track the evolution of superconductivity in the transition from TDBG to TBG. Finally, we discuss the contributions of different interactions to superconductivity and the implications of the results.

{\it Kohn-Luttinger-like mechanism.} We adopt a diagrammatic technique based on the KL theory~\cite{Kohn1965,Chubukov1993KL2D}. The diagrammatic method has previously given good agreement with experiments on both twisted~\cite{cea21Coulomb,phong2021band}, and non-twisted systems~\cite{Ghazaryan2021, cea2022superconductivity, JimenoPozo2023, DCL22, Dong2023, wagner23superconductivity,ZiyanLi2023, Pantaleon2023ReviewSC}. In this mechanism the pairing potential for Cooper pairs is the Coulomb interaction, screened by electron-hole excitations, longitudinal acoustic (LA) phonons~\cite{Kohn1965} and plasmons. We apply the RPA to compute the screened potential~\cite{cea21Coulomb}. We first consider the effect of electronic screening by calculating the bare electronic susceptibility, including  Umklapp processes. The susceptibility matrix reads,
\begin{align}
\mathcal{X}_0 (\mathbf{q})&= \frac{4}{V_{mBZ}}\sum_{k,m,n}\frac{f(\epsilon_{n,\mathbf{k}})-f(\epsilon_{m,\mathbf{k+q}})}{\epsilon_{n,\mathbf{k}}-\epsilon_{m,\mathbf{k+q}}} \times \nonumber \\
   &\times \langle \Psi_{n,\mathbf{k}}||\Psi_{m,\mathbf{k+q}}\rangle \otimes
    \langle \Psi_{m,\mathbf{k}}||\Psi_{n,\mathbf{k+q}}\rangle , 
\label{eq:1}
\end{align}
where $V_{mBZ}$ is the area of the Brillouin zone, $m$ and $n$ are band indexes and the factor 4 takes into account the spin and valley degeneracies. The function $f$ is the Fermi-Dirac distribution and $\epsilon_{n,\mathbf{k}}=E_{n,\vb{k}}-\mu$, where $\mu$ is the Fermi energy and $E_{n,\vb{k}}$ is the energy of the $n$-th band with momentum $\vb{k}$. We introduce a generalized form factor, $\langle \Psi_{n,\mathbf{k}}||\Psi_{m,\mathbf{k+q}}\rangle = \langle \Psi_{n,\mathbf{k}}| e^{i \vb{G} \cdot \vb{r}}|\Psi_{m,\mathbf{k+q}}\rangle$,
(see Sec.~\textcolor{red}{X} in Ref.~\cite{SM} for details). $\vb{G}_i$ is a reciprocal lattice vector resulting from the linear combination $\vb{G} = p\vb{G}_1+q\vb{G}_2$ with $p$ and $q$ integers and $\vb{G}_{1,2}$ the primitive reciprocal lattice vectors. This generalized matrix form of the susceptibility involves the connection matrix, $M(\vb{G})$~\cite{Bernevig2021I}, which enables a natural and computationally efficient treatment of the Umklapp processes. These processes let electrons interact with large momentum exchanges. The electronic susceptibility of a non-twisted system emerges as  $\lim_{\vb{G} \to 0} M(\vb{G}) = I$ of the twisted case, where the generalized form factor $\langle \Psi_{n,\mathbf{k}}||\Psi_{m,\mathbf{k+q}}\rangle$ that we use to describe moir\'e system is transformed to the usual $ \langle \Psi_{n,\mathbf{k}}|\Psi_{m,\mathbf{k+q}}\rangle$ that appears in non-twisted systems~\cite{cea2022superconductivity, JimenoPozo2023, wagner23superconductivity,ZiyanLi2023}. 

The phonon dressing of the electron-electron interaction is considered by computing the screened electronic susceptibility with RPA, renormalized by the interactions mediated by LA phonons,
\begin{equation}
    \label{eq:suscep_screened_PH}
    \chi_{\text{ph}}(\vb{q}) = \chi_{0}(\vb{q})\left(I - \mathcal{V}_{ph}(\vb{q},\omega)\chi_{0}(\vb{q})\right)^{-1}.
\end{equation} 
Here $\mathcal{V}_{ph}$ is the potential of the electron-phonon interaction (see Sec. \textcolor{red}{X} in Ref.~\cite{SM}). The screened electronic susceptibility is then used to obtain the screened Coulomb potential, $ \mathcal{V}_{scr}(\vb{q})$, within the RPA as
\begin{equation}
    \label{eq:screened_pot_final}
    \mathcal{V}_{scr}(\vb{q}) = \mathcal{V}_{C}(\vb{q})\left(I - \mathcal{V}_{C}(\vb{q})\chi_{ph}(\mathbf{q})\right)^{-1}.
\end{equation}
The screened potential defined here has a matrix structure, defined by reciprocal lattice vectors, $\vb{G} , \vb{G}'$. Hence, a simple effective potential in real space, $ \mathcal{V}_{scr}(\vb{r})$  cannot be directly obtained from the Fourier transform of Eq.~(\ref{eq:screened_pot_final}). We show in~\cite{SM} an approximate $ \mathcal{V}_{scr}(\vb{r})$ obtained by projecting the matrix implicit in Eq.~(\ref{eq:screened_pot_final}) onto  gaussian wavefunctions defined in the real space unit cell, see~\cite{Song22HeavyFermion}.

Finally, using this pairing potential, we analyze the emergence of the superconducting instability by self-consistently solving the linearized gap equation. The set of parameters used for the calculations is shown in Table.~\ref{tab:parameters}. We note that we have used the same set of parameters to determine the SC critical temperature in TBG~\cite{cea21Coulomb,Cea2023Superconductivity}, symmetric TTG~\cite{phong2021band} and non twisted graphene multilayers~\cite{JimenoPozo2023,ZiyanLi2023,cea2022superconductivity,Cea2023Superconductivity}. 

\begin{figure*}[ht]
    \centering
    \includegraphics[width = \textwidth]{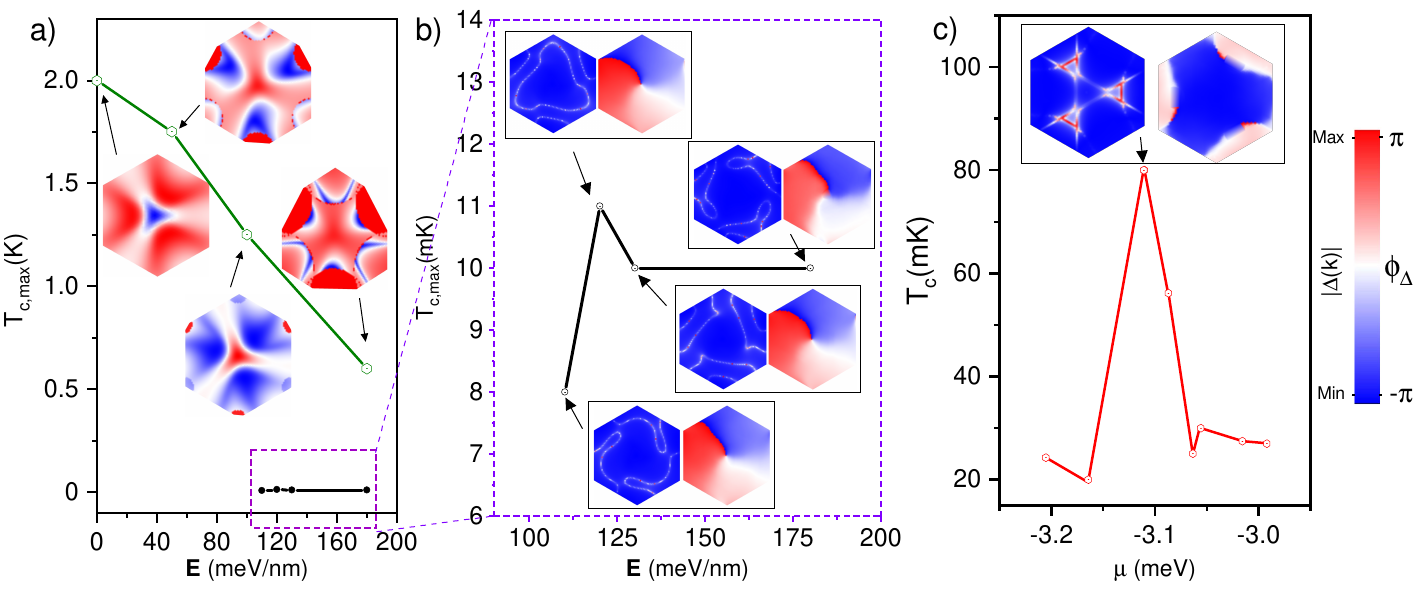}
    \caption{Maximum critical temperature $T_{c, max}$ and order parameter in different moiré stacks. For a) TBG at $1.1^{\circ}$ and b) TDBG at $1.1^{\circ}$, we display the evolution of the $T_{c, max}$ and the corresponding order parameter as a function of $\mathbf{E}$. In c) we show the SC dome for hTTG at $1.9^{\circ}$. The color density plots in all panels represent the norm $\abs{\Delta(k)}$ and complex phase $\phi_\Delta$, respectively, of the order parameter projected onto the valence band. In TBG the order parameter is real. For TDBG in b) and hTTG in c) the order parameters are complex; their amplitude (left) and phase (right) are plotted within the black rectangles.}   
    \label{fig:2}
\end{figure*}

{\it Superconductivity in TBG, TDBG and hTTG.} 
In order to study the similarities and differences between twisted bilayer graphene (TBG), twisted double bilayer graphene (TDBG), and helical trilayer graphene (hTTG), we define a dimensionless parameter, $0 \le \kappa_{h} \le 1$ which interpolates between TBG ($\kappa_{h} =0$) and TDBG and hTTG ($\kappa = 1).$ As sketched in Figure~\ref{fig:1}, the parameter $\kappa$ increases gradually the coupling between the two outer and the two inner layers in TDBG, or between one of the outer layers and the other two, in hTTG. Details of  the band structure of TDBG and hTTG, as well as the modifications introduced by the Hartree potential are given in~\cite{SM}.

Figure~\ref{fig:1} shows the real space configuration of a) the transition from TDBG ($\kappa_{h}=1$) to effective TBG+2 MG ($\kappa_{h}=0$), and b) the transition from hTTG ($\kappa_{h}=1$) to effective TBG+1 MG ($\kappa_{h}=0$), where $\kappa_{h}\in [0,1]$ refers to the coupling scaling parameter. This parameter allows us to perform a continuous transition from the TDBG or hTTG structure to the TBG configuration, enabling the tracking of electronic properties along the transition. In TDBG, the coupling scaling parameter $\kappa_{h}$ modulates the strength of interlayer hoppings between adjacent layers within the BBG systems, which are rotated in opposite directions to form the TDBG system. When $\kappa_{h}$ is set to zero, the only remaining interlayer hoppings in the system are those corresponding to the central rotated layers, resulting in an effective TBG plus two isolated MG. In Fig.~\ref{fig:1}c) we show the low energy band structure of TDBG ($\kappa_{h}=1$, in solid blue) and effective TBG ($\kappa_{h}=0$, in dashed gray) in the absence of external potentials. In hTTG, the layers are rotated consecutively with the same twist angle, in an staircase configuration~\cite{Guerci2023Chern} and the parameter $\kappa_{h}$ adjusts the interlayer hopping between two adjacent layers (e.g. middle and top layers), 
such that if $\kappa_{h}=0$ the system is an effective TBG plus an isolated MG, as illustrated in Fig.~\ref{fig:1}b). The low energy band structure of hTTG ($\kappa_{h}=1$, in solid blue) and effective TBG ($\kappa_{h}=0$, in dashed gray) are shown in Fig.~\ref{fig:1}d) and e) for two different twist angles as indicated.

In Fig.~\ref{fig:2} we present the superconducting critical temperatures and order parameters (OPs) arising from the screened Coulomb interaction for TBG, TDBG and hTTG. We set a twist angle $\theta=1.1^{\circ}$ for the TDBG and TBG and $\theta=1.9^{\circ}$ for hTTG. We also study hTTG at $1.1^{\circ}$ in Ref.~\cite{SM}. In our parameterization, $\theta=1.1^{\circ}$ and $\theta=1.9^{\circ}$ are the TBG~\cite{Koshino2018a} and hTTG~\cite{Foo2023TTG,Yang2023Multi} magic angles, respectively. Figure~\ref{fig:2}a) and b) illustrate the critical temperature dependence on the external electric field, $\mathbf{E}$, for TBG (a) and TDBG (b). In TBG, the external field shifts in opposite ways the two Dirac cones defined in the two central marrow bands, although it leads to a reduction and smearing of the DOS peaks~\cite{Koshino2014Optical}, as shown in Fig.~S4a). In TDBG the electric field opens a gap and flattens the bands, as shown in Fig.~S4b). For TDBG, we observe a nearly constant critical temperature of approximately $T_{c}\approx 10$ mK within the range of electric fields spanning from 110 to 180 meV/nm. This finding aligns well with the experimentally established order of magnitude for the critical temperature~\cite{Su2023Superconductivity}. For TBG, we find a maximum critical temperature of $T_{c}\approx 2$ K in absence of an external field, in good agreement with experiments and previous calculations~\cite{cea21Coulomb,Cea2023Superconductivity}. We notice that the critical temperature of TBG is notably reduced by the inclusion of an external field, dropping to $T_{c}\approx 0.6$ K at $\mathbf{E}=180$ meV/nm, also in line with recent experimental measurements~\cite{Dutta2024}. We attribute this effect to a smearing of the vHs and the electric field induced wavefunction heterogeneity (see Secs. \textcolor{red}{IX} and \textcolor{red}{VI} in Ref.~\cite{SM}). The case of hTTG at $\theta=1.9^\circ$ is shown in Fig.~\ref{fig:2}c). We find a maximum critical temperature of $T_{c}\approx 80$ mK for a Fermi energy close to the vHs. The critical temperature quickly vanishes when the Fermi energy moves a few meV away from the vHs, following a similar tendency as the density of states. In the transition to TBG, by reducing the coupling strength, the vHs splits and the SC is immediately suppressed (see Sec. \textcolor{red}{VI} in Ref.~\cite{SM}). This indicates that the SC state in hTTG is fragile and sensitively depends on the twist angle between graphene layers. 

\begin{figure*}[ht]
    \centering
    \includegraphics[width = 0.8\textwidth]{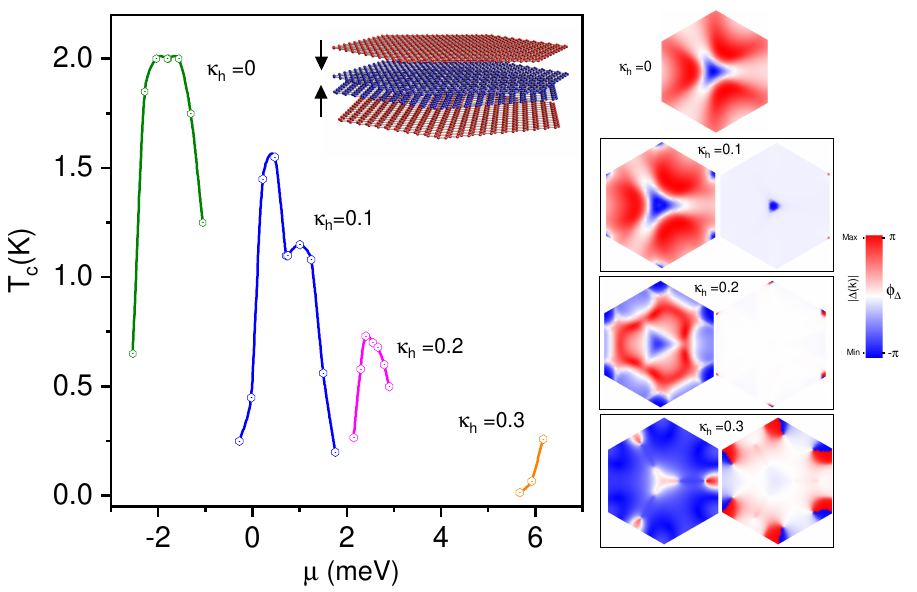}
    \caption{Superconducting domes and order parameters along a transition from TBG ($\kappa_h=0$) to TDBG with slightly decoupled outer layers ($\kappa_h=0.3$), as shown in the inserted lattice structure. 
    The order parameters are calculated at the Fermi energies which gives the maximum $T_c$ within each dome. $\mathbf{E}=0$ in all cases. The color density plots represent the norm $\abs{\Delta(k)}$ and complex phase $\phi_\Delta$, respectively, of the order parameter projected onto the valence band. Density plots of the complex order parameters are within black rectangles including the amplitude (left) and the phase (right). For $\kappa =0 $ (TBG) the order parameter is real.}   
    \label{fig:3}
\end{figure*}

In Fig.~\ref{fig:3} we show the critical temperature as a function of Fermi energy, along the TBG-TDBG transition for different values of the coupling scaling factor $\kappa_{h}$, at $\mathbf{E}=0$. The shrinking of the superconducting dome for increasing values of the coupling factor makes it clear that the interlayer hoppings within the non-twisted bilayers are detrimental to superconductivity, and no SC is found for TDBG without electric field at $\kappa_{h}=1$. As the coupling with the outer layers increases, the Hartree strength is suppressed (see Sec.\textcolor{red}{VII} in Ref.~\cite{SM}). We attribute this effect to a charge redistribution to the outer layers, consistent with the absence of a superconducting phase in other twisted multilayer systems with a similar configuration~\cite{Riffo2024Behavior}. 

The insets in Figure~\ref{fig:2} show the superconducting order parameter (OP), $\Delta(\vb{k})$, in the moiré Brillouin zone of these materials. In Fig.~\ref{fig:2}a) the OP of the TBG is real-valued for every external field considered. For zero and 50 meV/nm external fields the OP changes sign in the vicinity of the $\Gamma$-point, however, as the external field increases the OP maintains a constant sign in the moiré Brillouin Zone as in the case of $\vb{E}=$100 meV/nm and 180 meV/nm. Despite the change of sign these OP cannot be considered nodal since their average is far from being zero. These OPs seem to be non-zero in the full moiré Brillouin zone. It is worth noting that in this case the Fermi energy is very close to half-filling and therefore the vicinity of the Fermi surface occupies a significant fraction of the moiré Brillouin Zone. To determine the relative sign of the OP in the two valleys it is necessary to include an intervalley interaction that splits the valley degeneracy as it was done in~\cite{JimenoPozo2023}, which is out of the scope of this paper. In contrast to TBG, the OP of TDBG, shown in the insets of Fig.~\ref{fig:2}b), does not change qualitatively upon varying the external electric field. These OPs are non-zero only along the contour of the Fermi surface of TDBG, similarly to those of non-twisted graphene stacks~\cite{JimenoPozo2023, wagner23superconductivity,ZiyanLi2023,Pantaleon2023ReviewSC}, consistent with a critical temperature much lower than the bandwidth. The low vallue of the critical temperature implies as well that the results shown in Fig.~\ref{fig:2}b) for TDBG and hTTG are moe sensitive to numerical unnacuracies than the results for TBG. We expect an uniform complex phase that approaches either $\pi$ or $-\pi$ as the grid size increases and therefore we consider the order parameter symmetry of TDBG to be intravalley $s$-wave, similar to that observed in TBG and hTTG. The OP for hTTG at $\theta=1.9^{\circ}$ is shown in the insets of Fig.~\ref{fig:2}c). In the insets of Fig.~\ref{fig:3} we show the evolution of the OP from the effective TBG to a slightly coupled TDBG with $\kappa_{h}=0.3$. 

\onecolumngrid

\begin{table}[ht]
\centering 
\begin{tabular}{|c|c|c|c|c|c|c|c|c|c|c|c|c|} 
\hline
\multicolumn{8}{|c|}{Electronic bands}                                                                                                                                                 & \multicolumn{2}{c|}{Coulomb potential}     & \multicolumn{3}{c|}{\begin{tabular}[c]{@{}c@{}}Deformation potential\\and elastic constants\end{tabular}}  \\ 
\hline\hline
$a$(\r{A}) & $v_{F}$ (eV\r{A}) & $\gamma_{1}$ (eV) & $\gamma_{3}$ (eV) & $\gamma_{4}$ (eV) & $\Delta'$ (eV) & $g_{1}$ (eV) & $g_{2}$ (eV) & $d_{g}$ (nm) & $\epsilon$ & $D_{0}$(eV) & $\lambda$ (eV\r{A}$^{-2}$) & $\mu$ (eV\r{A}$^{-2}$) \\ 
\hline
2.46  & 5.253        & 0.4                              & 0.32                             & 0.044                            & 0.05                     & 0.0797       & 0.0975       & 40         & 10                          & 20           & 3.25                       & 9.44                                                           \\
\hline
\end{tabular}
\caption{Parameters employed in the calculations. They fully define the continuum models and the electron-electron/electron-phonon potentials considered in this work. Note that $\gamma_{1,3,4}$ and $\Delta^{\prime}$ are not included in the model of hTTG.}
\label{tab:parameters}
\end{table}


\twocolumngrid

\textit{Contributions to superconductivity.} Next, we consider the physical processes which drive superconductivity. For TDBG with an $\mathbf{E}$ field, we can only resolve $T_{c}$ when both Umklapps and phonon dressing are present, and we find $T_{c}\approx 10$ mK, comparable to the experiment~\cite{Su2023Superconductivity} (see  Fig. \textcolor{red}{S13} in Ref.~\cite{SM}). If either factor is absent $T_c$ drops to below 1 mK. Similarly, in TBG we find a critical temperature below 1 mK when either Umklapps or phonons are omitted. However, the inclusion of both leads to a very significant increase in $T_c$ to 2 K, indicating that Umklapp phonon dressing of the Coulomb interaction is crucial for superconductivity in the material, distinguishing it from other arrangements. In stark contrast, for hTTG, including Umklapps and phonons yields a $T_c$ of 80 mK, but $T_c$ only diminishes to 60 mK if Umklapps are omitted. However, phonon dressing remains a key factor for superconductivity in this material, as without it, superconductivity drops below 1 mK. Therefore, all of these moiré stacks differ from non-twisted graphene, for which electron-electron interactions suffice to give rise to critical temperatures in agreement with experiments, while phonon dressing and Umklapp processes do not play a major role~\cite{ZiyanLi2023,Pantaleon2023ReviewSC,JimenoPozo2023}. In addition, a brief analysis of the homogeneity of the wave functions for the three considered systems is shown in \cite{SM}.   

The fact that the superconducting critical temperature in TBG is about two orders of magnitude higher than in TDBG and hTTG correlates well with the significantly larger effect of the electron-electron interaction on the shape and width of the electronic bands~\cite{Guinea2018Electrostatic}, although the parameters which describe the bare interactions are the same in all systems, see Table~\ref{tab:parameters}. The relevance of Umklapp processes  can be traced back to the existence of significant inhomogeneities in the charge distribution at scales of about 1/3 of the moiré scale. This is the regime well described by the topological heavy fermion model~\cite{Song22HeavyFermion}.

{\it Discussion.} Superconductivity is ubiquitous in graphene multilayers. However, its strength, as well as the conditions for its appearance and enhancement show marked contrasts between different arrangements. The discovery of superconductivity in twisted double bilayer graphene~\cite{Su2023Superconductivity}, after sustained efforts~\cite{Burg2019Correlated,Shen2020Correlated,Cao2020SpinPolarized,Liu2020SpinCorrelated,He2020SymBreak,Kuiri2022Spontaneous}, is a valuable addition to the field, while superconductivity has yet not been reported in helical trilayer graphene.

We have investigated superconductivity in TBG, TDBG and hTTG mediated by the screened long-range Coulomb interaction in the RPA. Our model incorporates electron-electron and electron-phonon interactions, We pay special attention to Umklapp processes, induced by the existence of significant inhomogeneities within the unit cell and in the moiré Brillouin Zone. 

A one parameter interpolation scheme allows us to track the evolution of the superconducting phase from TBG to TDBG, and to hTTG. We find critical temperatures in good agreement with experiments on both TDBG and TBG. We also predict weak superconductivity in hTTG with $T_c=80$~mK, and a significant rise in $T_c$ in stacks where spin-orbit coupling is induced by the proximity to few-layers of WSe$_2$ \cite{Arora2020_TBG_WSe2,zhang2022spin,Zhang2022_promotionSC,holleis2023SC,Su2023Superconductivity}. 

Superconductivity in all the twisted systems considered here is weakened when the calculations are performed starting from a model with two electronic flavors, instead of four, as shown in Sec. \textcolor{red}{XII} in Ref.~\cite{SM}, in contrast to the enhancement found for the non-twisted stacks~\cite{JimenoPozo2023,ZiyanLi2023}. The decrease of superconductivity due to the reduction of flavor symmetry has been observed in a recent experiment on TBG~\cite{Dutta2024}. The role played by the Ising spin-orbit coupling and other effects induced by WSe$_2$ in stabilizing superconductivity versus other correlated phases are still unclear and require further investigation both theoretically and experimentally~\cite{Yang2023Revived,chou2024topological}. It is worth noting that this calculation roughly estimates the effect of full spin or valley polarization. The fact that $T_c$ in TBG is changed from 2K to 1K implies that flavor polarization does not substantially suppress superconductivity~\cite{goldstone}.

It is finally worth noting that isolated theoretical estimates of the critical temperature of a superconductor typically have a strong dependence on the choice of parameters used in the calculation. On the other hand, general trends calculated using the same set of parameters, can be expected to be more reliable. Our calculations describe semi-quantitatively systems where the superconducting properties vary over a large range. This fact suggests that the interactions studied here play an significant role in the superconductivity of these materials.

{\it \bf Acknowledgments.} We thank Zhen Zhan, Saul Herrera, Gerardo Naumis and Tommaso Cea for discussions. M.L. expresses gratitude to all members of the Theoretical Modelling group at IMDEA Nanoscience for their warm hospitality during his research stay.  We acknowledge support from the Severo Ochoa programme for centres of excellence in R\&D (CEX2020-001039-S / AEI / 10.13039/501100011033, Ministerio de Ciencia e Innovaci\'on, Spain);  from the grant (MAD2D-CM)-MRR MATERIALES AVANZADOS-IMDEA-NC, NOVMOMAT, Grant PID2022-142162NB-I00 funded by MCIN/AEI/ 10.13039/501100011033 and by “ERDF A way of making Europe”. M.L. acknowledges support from the Research Grants Council (RGC) of Hong Kong Special Administrative Region of China (Project Nos. 17301721, AoE/P-701/20, 17309822, HKU C7037-22GF), and the HKU Seed Funding for Strategic Interdisciplinary Research. We thank the HPC2021
system under the Information Technology Services and the Blackbody HPC system at the Department of Physics, University of Hong Kong, as well as the Beijng PARATERA Tech CO.,Ltd. (URL: https://cloud.paratera.com) for providing HPC resources that have contributed to the research results reported within this paper.

\bibliographystyle{apsrev4-2} 

%

\clearpage
\onecolumngrid

\begin{center}
 {\Large {\it Supplementary information for:} \\
 \vspace{0.15cm} 
  Evolution of Superconductivity in Twisted Graphene Multilayers} \\
 \vspace{0.25cm}
{\normalsize Min Long, Alejandro Jimeno-Pozo, H\'ector Sainz-Cruz, Pierre A. Pantale\'on\footnote{pierre.pantaleon@imdea.org} and Francisco Guinea 
 }
 
\end{center}

\setcounter{equation}{0}
\setcounter{figure}{0}
\setcounter{table}{0}
\setcounter{page}{1}
\makeatletter
\renewcommand{\theequation}{S\arabic{equation}}
\renewcommand{\thefigure}{S\arabic{figure}}
\setcounter{secnumdepth}{1}

\tableofcontents

\section{Continuum model for TDBG}
\label{sec:ContinuumTDBG}

\begin{figure*}[ht!]
    \centering
    \includegraphics[scale=0.3]{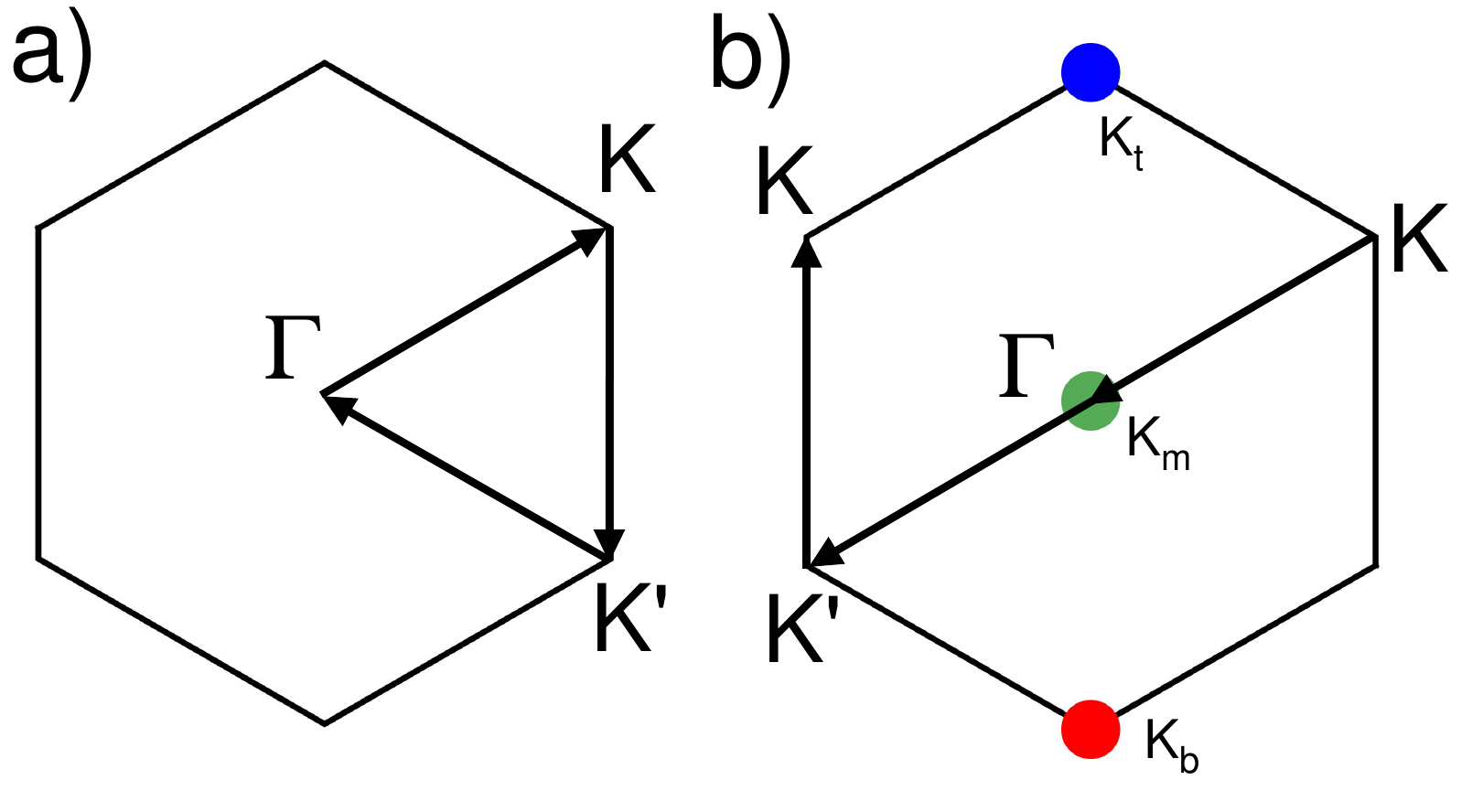}
    \caption{Moiré Brillouin zone for a) TDBG and b) hTTG with the corresponding high symmetry points. In b) we show the position of the Dirac cones of the trilayer configuration. Arrows in both figures indicate the path used to calculate the band structures.}   
    \label{fig:mbz}
\end{figure*}

The relative twist between two Bernal stacks of bilayer graphene leads to the appearance of a moir\'e pattern. The size of the supercell, $L_m=a/2\sin\left(\theta/2\right)$, dramatically increases with the twist angle, $\theta$, $a=0.246$ nm being the lattice constant of graphene. We describe the low-energy band structure of the TDBG within the continuum model introduced in the Refs.~\cite{LopesDosSantos2007b,Bistritzer2011} for the case of the TBG, and generalized in Refs.~\cite{Koshino2019a,Chebrolu2019a}. This model is meaningful for sufficiently small angles, so that an approximately commensurate structure can be defined for any twist. The mBZ shown in Fig.\ref{fig:mbz}a), resulting from the folding of the two BZs of each bilayer, has the two reciprocal lattice vectors:
\begin{equation}
\vb{G}_1= \frac{2\pi}{L_{m}}(1/\sqrt{3},1)\text{, } \vb{G}_2=\frac{4\pi}{L_{m}}(-1/\sqrt{3},0).
\end{equation}

For small twists, the coupling between the two valleys at $\vb{K}=(4\pi/3a)(1,0)$ and $-\vb{K}$ of the unrotated bilayer can be safely neglected, as the interlayer hopping has a long wavelength modulation. Then, for the sake of simplicity, in what follows we focus only on the $K$-valleys of each bilayer. The Hamiltonian of the TDBG is represented by the $8\times 8$ matrix:
\begin{equation}
H_{TDBG}=\begin{pmatrix}
H_0(\vb{k}_1)&H_1(\vb{k}_1)&0&0&&\\
H_1^\dagger (\vb{k}_1) &H'_0(\vb{k}_1)&U(r)&0&\\
0&U(r)^\dagger& H_0(\vb{k}_2)&H_1(\vb{k}_2)\\
0&0&H_1^\dagger (\vb{k}_2) &H'_0(\vb{k}_2)
\end{pmatrix}
\label{eq: HamilTDBG}
\end{equation}
acting on the Nambu spinor $\Psi^T=\left(\psi_{A1},\psi_{B1},\psi_{A2},\psi_{B2},\psi_{A3},\psi_{B3},\psi_{A4},\psi_{B4}\right)$, whose entries are labelled by the sub-lattice ($A/B$) and layer ($1,\dots 4$) indices. Here we defined $\vb{k}_l=R\left((-)^{l-1}\theta/2\right)\left( \vb{k}-\vb{K}_l \right)$, $R(\theta)$ being the $2\times 2$ matrix describing the counter-clock-wise rotation of the angle $\theta$, and:
\begin{align}
H_0(\vb{k})&=\begin{pmatrix}
    0&\hbar v_F k_-\\
    \hbar v_F k_+& \kappa_{h}\Delta^{\prime}
    \end{pmatrix} \notag \\
H'_0(\vb{k})&=\begin{pmatrix}
   \kappa_{h} \Delta^{\prime}&\hbar v_F k_-\\
    \hbar v_F k_+& 0
    \end{pmatrix} \\
H_1(\vb{k})&=\kappa_{h}
    \begin{pmatrix}
   -\hbar v_4k_-&-\hbar v_3k_+\\   \notag
    \gamma_1& -\hbar v_4 k_-
    \end{pmatrix}
\end{align}
where $k_{\pm}=k_x\pm ik_y$, $v_i=(\sqrt{3}/2)\gamma_i a/\hbar$, $\gamma_1=0.4$ eV, $\gamma_3=0.32$ eV, $\gamma_4=0.044$eV, $\Delta^{\prime}=0.05$eV, $\hbar v_F/a=2.1354$eV 
\cite{Koshino2019a}. $\kappa_{h}\in[0,1]$ is a parameter to continuously convert TDBG to TBG by reducing the hopping parameters in the two BBGs that form the TDBG. The $2\times 2$ matrix $U$ describes the moir\'e potential generated by the hopping amplitude between $p_z$ orbitals localized at opposite layers of the two twisted surfaces. In real space, $U(\vb{r})$ is a periodic function in the moir\'e unit cell. In the limit of small angles, its leading harmonic expansion is determined by only three reciprocal lattice vectors~\cite{LopesDosSantos2007b}:
$U(\vb{r})=U(0)+U\left(-\vb{G}_1\right)e^{-i\vb{G}_1\cdot\vb{r}}+
U\left(-\vb{G}_1-\vb{G}_2\right)e^{-i\left(\vb{G}_1+\vb{G}_2\right)\cdot\vb{r}}$,
where the amplitudes $U\left(\vb{G}\right)$ are given by:

\begin{align}
U\left(\boldsymbol{0}\right) & =\left(\begin{array}{cc}
g_{1} & g_{2}\\
g_{2} & g_{1}
\end{array}\right),\nonumber \\
U\left(-\boldsymbol{G_{1}}\right) & =\left(\begin{array}{cc}
g_{1} & g_{2}e^{-\frac{2\pi i}{3}}\\
g_{2}e^{\frac{2\pi i}{3}} & g_{1}
\end{array}\right),\label{eq: Ur}\\
U\left(-\boldsymbol{G_{1}}-\boldsymbol{G_{2}}\right) & =U\left(-\boldsymbol{G_{1}}\right)^{\ast},\nonumber 
\end{align}
In the following we adopt the parametrization of the TBG given in the Ref.~\cite{Koshino2018a}: $g_1=0.0797$ eV and $g_2=0.0975$ eV. The difference between $g_1$ and $g_2$ accounts for the corrugation effects where the interlayer distance is minimum at the $AB/BA$ spots and maximum at $AA$ ones, or can be seen as an effective model of a more complete treatment of lattice relaxation \cite{Guinea2019a}. The difference in electrostatic energy between the adjacent bands is given by~\cite{Koshino2019a}  

\begin{equation}
V_{E}=\frac{1}{2}\Delta V\left(\begin{array}{cccc}
3I_{2} & 0 & 0 & 0\\
0 & I_{2} & 0 & 0\\
0 & 0 & -I_{2} & 0\\
0 & 0 & 0 & -3I_{2}
\end{array}\right),
\label{eq: TBGVias}
\end{equation}
where $I_{2}$ is a $2\times 2$ identity matrix and $\Delta V$ an (out of plane) external electrostatic potential. The transition from a TBG to a TDBG is introduced by a coupling parameter $\kappa_{h}$, such that $\kappa_{h}=0$ if the external graphene layers are decoupled from the internal TBG structure and $\kappa_{h}=1$ for the full TDBG.

\section{Continuum model for \texorpdfstring{\lowercase{h}{T}} TTG}
\label{sec:ContinuumTTG}
We now consider the transition from TBG to hTTG. Labeling the layers consecutively, we twist layers bottom ($b$), medium ($m$) and top ($t$)  by $-\theta$, $0$ and $\theta$ about a fixed hexagon center. To describe the low-energy physics at small angles, we adopt a valley-projected continuum Hamiltonian~\cite{Foo2023TTG,Mora2019Flatbands} which is the simplest case of the multilayer case described in Ref.~\cite{Cea2019Twists} and is given by
\begin{equation}
\begin{split}
H  &= \begin{pmatrix}
H_b \left(\mathbf{q_b} \right) &  U_{bm} (\mathbf{r}-T_{bm}) & 0\\
U_{bm}^\dagger (\mathbf{r}-T_{bm}) & H_m \left(\mathbf{q_m} \right) &  \kappa_{h} U_{mt} (\mathbf{r}-T_{mt}) \\
0 & \kappa_{h} U_{mt}^\dagger (\mathbf{r}-T_{mt}) & H_t \left(\mathbf{q_t} \right)
\end{pmatrix}, \\
H_\ell&= - \hbar v_F \mathbf{q_\ell} \cdot \left( \sigma_x, \sigma_y \right), \\
\end{split}
\label{eq: Hamiltonian}
\end{equation}
where $\mathbf{q_\ell}=\left( \mathbf{k} - \mathbf{K}_\ell \right)$ with $\ell=\{m,b,t \}$ and
$U(r)$ given by Eq.~\ref{eq: Ur}. We utilize the same set of parameters as in the TDBG case. The positions of the Dirac cones in the rotated system are indicated in Fig.~\ref{fig:mbz}b). The parameter $\kappa_{h}\in[0,1]$ allows for a continuous transformation from hTTG to TBG by modulating the moiré potential strength between the middle and top layers. In the above equation, we introduce a shift vector $T$ to account for a local displacement between layers (for detailed information, please refer to Ref.~\cite{Foo2023TTG}). In this study, we find that, with our set of parameters, the narrowest bands are obtained by setting $T_{bm}=0$ and $T_{mt}=-(L_1+2L_2)/3$, corresponding to an ABA stack configuration. 

\section{Self-consistent Hartree interaction: Theory}
\label{sec:HartreeTheory}
The Hartree potential is introduced to the main Hamiltonian as a matrix $H_{Hartree}$ with elements given by~\citep{Cea2022Electrostatic,Pantaleon2021Narrow}, 
\begin{equation}
\rho_{H}(\boldsymbol{G})=4V_{C}\left(\boldsymbol{G}\right) \int \frac{d^{2}\boldsymbol{k}}{V_{mBZ}}\sum_{n,\beta,\vb{G'}}u^*_{\beta,n,\vb{G'+G}}(\vb{k})u_{\beta,n,\vb{G'}}(\vb{k}).
\label{eq: rho components} 
\end{equation}
where  $n$ is a  band index and $\beta$ refers to the layer and sub-lattice degrees of freedom. The quantity $V_{C}\left(\boldsymbol{G}\right)=2\pi e^{2}/\epsilon \left|\boldsymbol{G}\right|$ is the Fourier transform of the Coulomb potential evaluated at $\boldsymbol{G}$, where $\epsilon$ is the dielectric constant. The factor $4$ takes into account spin/valley degeneracy, and $V_{mBZ}$ is the area of the mBZ. $u_{n}$ is a self-consistent eigenvector resulting from the numerical diagonalization of the total Hamiltonian. In terms of the connection matrix it can be written as
\begin{equation}
\rho_{H}(\boldsymbol{G})=4V_{C}\left(\boldsymbol{G}\right) \int \frac{d^{2}\boldsymbol{k}}{V_{mBZ}}\sum_{n}\langle u_{n}(\vb{k})|M(\vb{G})|u_{n}(\vb{k})\rangle,
\label{eq: rho con M} 
\end{equation}

with
\begin{equation}
    M(\vb{G})_{ij} =    \begin{cases}
       \text{$I_{\beta}$} &\text{ if $\vb{G}_i-\vb{G}_j = \vb{G}$},\\
       \text{$0_{\beta}$} &\text{ if $\vb{G}_i-\vb{G}_j \neq \vb{G}$}
    \end{cases}
\end{equation}
The above definition connects the Fourier components of the eigenvector $u_{n}(\vb{k})$ that differ by a vector $\vb{G}$. $I_\beta$ is a $\beta$-dimensional identity matrix, with $\beta=8$ representing layer and sublattice degeneracy in the case of TDBG and $\beta=6$ for hTTG. We can express the Fourier expansion of the Hartree potential in real space as 
\begin{equation}
V_\text{H}(\vec{r})=V_{0}\sum_{j}\rho_\text{H}(\boldsymbol{G}_{j})e^{i\boldsymbol{G}_{j}\cdot\vec{r}},
\label{eq: VhRealSpace}
\end{equation}
with $\rho_\text{H}(\boldsymbol{G}_{j})$ in general a complex number and $V_{0}=e^{2}/\varepsilon L_m$ the effective Coulomb potential. To solve the self-consistent Hartree Hamiltonian, we approximate the charge distribution as $\rho_{H}=\overline{\rho}_{H}+\delta\rho_{H}$, where $\overline{\rho}_{H}$ is a constant that takes into account the total density from all bands not included in the calculations~\citep{Cea2022Electrostatic}. The charge distribution is fixed by considering a homogeneous state at the charge neutrality point (CNP). Therefore, the integral in Eq.~(\ref{eq: rho components}) is evaluated only over energy levels between CNP and the Fermi energy. 

It is worth noting that, for an arbitrary overlap ${\vb{k}+\vb{q}+\vb{G'}, \vb{k}+\vb{G}+\vb{G'}}$, the form factor defined with the connection matrix $\langle u_{n}(\vb{k}+\vb{q})|M(\vb{G})|u_{n}(\vb{k})\rangle$ for moiré systems is a generalization of the ordinary form factor $\braket{u_{n}(\vb{k}+\vb{q})}{u_{n}(\vb{k})}$, where $\vb{G} = \vb{0}$, and thus $M(\vb{G}) = I$. Moreover, we find that the connection matrix allows us to describe the symmetry operations on the Hamiltonian while calculating some physical quantities, such as the screened Coulomb potential, and is better for numerical implementation.

\begin{figure}
\includegraphics[scale=0.4]{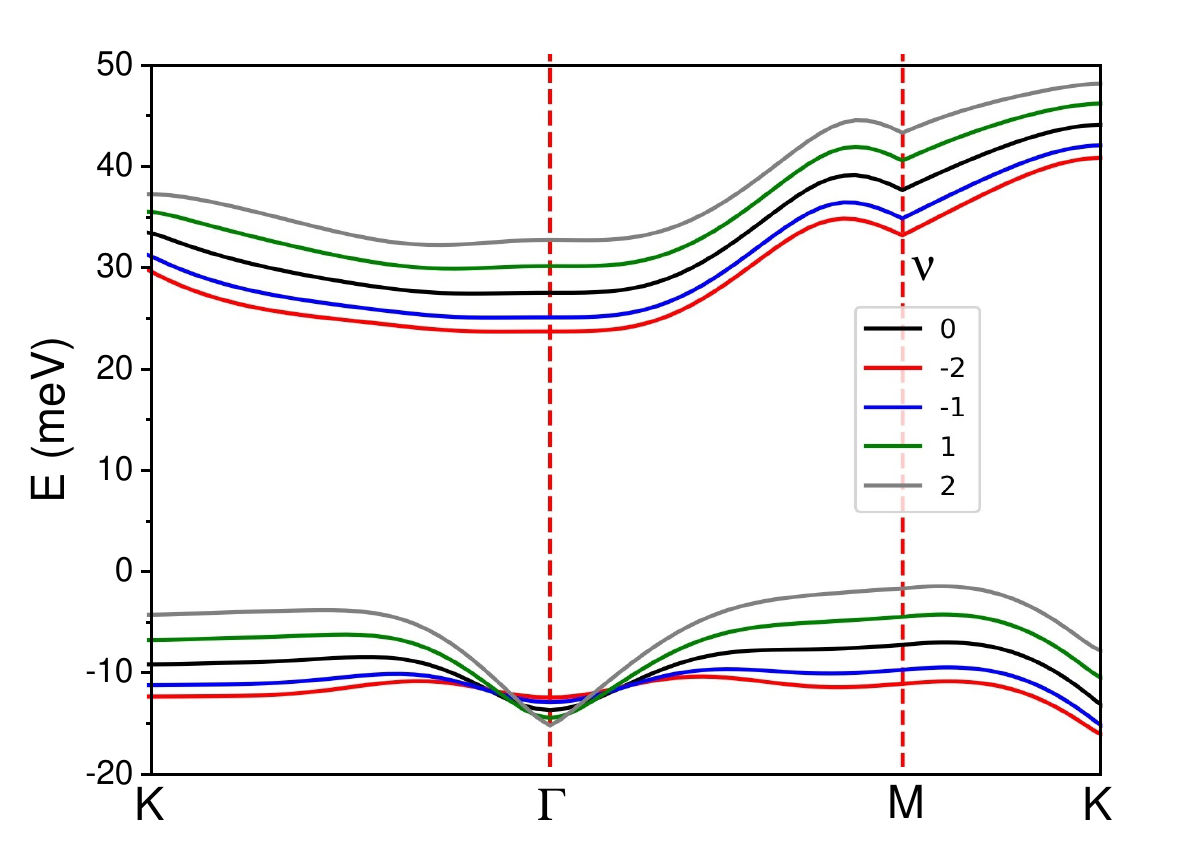}
\caption{Hartree band structure of TDBG. The two central narrow bands of are shown, as a function of filling, for $\theta=1.10^\circ$ and in the presence of the electric field $E = 0.05$eV/nm.}
\label{fig:HF_effect}
\end{figure}

\section{Self-consistent Hartree interaction: TDBG}

As described in Refs.~\cite{Pantaleon2021Narrow,Cheung2022Atomistic}, the different charge distribution within the mBZ in TDBG results in a small band distortion due to the Hartree potential, even when the bandwidth of the active band is smaller than the effective Coulomb interaction. Figure~\ref{fig:HF_effect} illustrates the impact of the Hartree interaction on the middle bands of TDBG. In the lower band, the regions around $\Gamma$ are nearly insensitive to the Hartree potential, while the regions at the mBZ edges undergo uniform shifts. In the upper band, the shift is uniform. Consequently, in our superconductivity (SC) calculations, we did not consider directly the Hartree effect but we tracked its strength as a function of the coupling parameter, as shown in Fig.~\ref{fig:HartreeTBG} and Fig.~\ref{fig:HartreeTTG1}. It is worth noting that the strength of the Hartree potential is directly related to the strength of the Umklapp processes through the form factor, $\langle u_{n}(\vb{k})|M(\vb{G})|u_{n}(\vb{k})\rangle$. We have found that the strength of this factor also impacts the screening potential and related quantities. The form factors in Eq. \textcolor{red}{1}, Eq.~\ref{eq: rho con M}, and Eq.~\ref{eq: GapEquation} can be directly correlated with the strength of the Hartree potential. In particular, the susceptibility is large if the Fermi energy is at or close to a van Hove singularity (vHS), but its magnitude is even larger if the form factor also contributes.

\begin{figure}
\includegraphics[width = \textwidth]{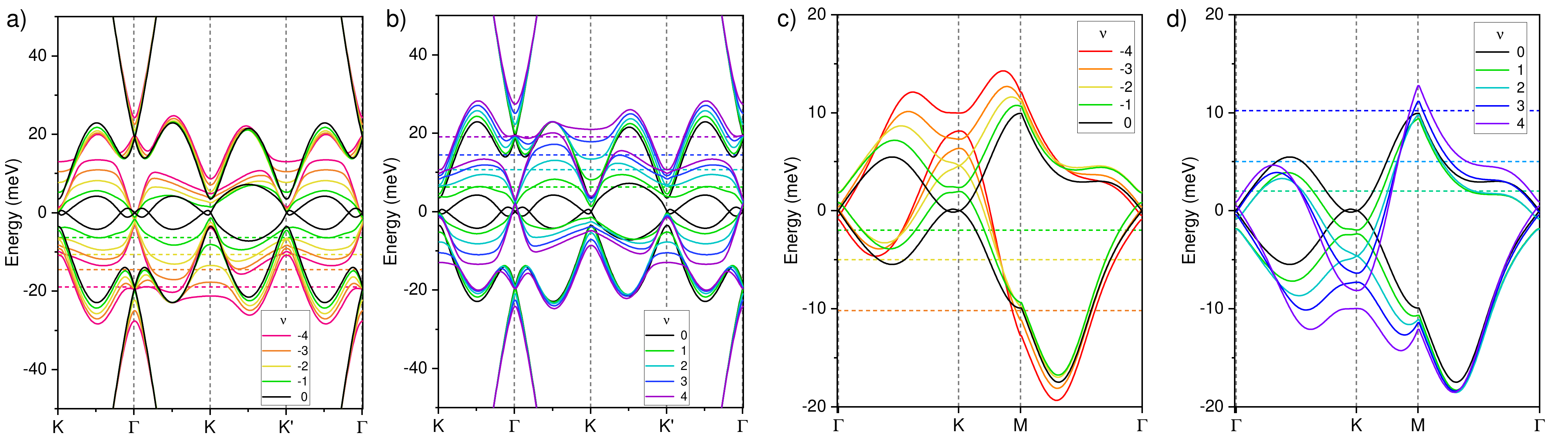}
\caption{Hartree band structure of hTTG. Filling and color scheme is indicated in each panel. Twist angle is set to $\theta=1.1^\circ$ in a) and b) and $\theta=1.9^\circ$ in c) and d). Horizontal colored dashed lines are the positions of the Fermi energies for the corresponding filling.}
\label{fig:HF_effectTTG}
\end{figure}

\section{Self-consistent Hartree interaction: \texorpdfstring{\lowercase{h}{T}} TTG}

In the narrow bands of TBG or slightly coupled TDBG (see Fig.~\ref{fig:HartreeTBG}f) and g)), the charge density is distributed in a ring around the AA centers at $\Gamma$ and precisely at the AA centers at the mBZ edges~\cite{San-Jose2012NonAbelian,Guinea2018Electrostatic,Rademaker2019Charge, HungNguyen2021,NavarroLabastida2022Why}. The filling-dependent TBG Hartree potential is triangular and centers around the AA sites~\cite{Guinea2018Electrostatic,Rademaker2019Charge}. Consequently, while the states around $\Gamma$ remain insensitive to the filling, the states at the edges of the mBZ undergo uniform shifts. This mechanism underlies the pinning of the van Hove singularity (vHS) with the Fermi energy in TBG~\cite{Cea2019Pinning}, alternated TTG~\cite{Phong2021BandTTG} and even in supermoiré TTG structures~\cite{Hao2024Robust}. In the case of hTTG, as depicted in Fig.~\ref{fig:HF_effectTTG}, for $\theta=1.1^\circ$ in a) and b), the Hartree potential does not significantly alter the bands around $\Gamma$ and $K$, but it introduces small modifications around $K'$. In other regions, the distortion is strong, a similar effect is found for $\theta=1.9^\circ$ in c) and d). Therefore, the behaviour of the hTTG system is in someway similar to TDBG where the charge distribution is different along different points within the mBZ~\cite{Pantaleon2021Narrow,Cheung2022Atomistic}. This effect was also found to be the source of the reduction of the Hartree potential in TDBG~\cite{Cheung2022Atomistic}, bilayer graphene (BG/hBN) or rhombohedral trilayer graphene (RTG/hBN) with an hBN substrate~\cite{Pantaleon2021Narrow}. In addition, as mentioned before, we also found that the reduction of the Hartree strength is a consequence of small contributions from the form factor which is a measure of the strength of the Umklapp processes. Thus, as in the case of TDBG the obtained critical temperature is marginal with respect to TBG and the other alternating-twist stacks. We expect similar effects in BG/hBN, RTG/hBN and other graphitic moiré systems with non-uniform charge distribution. However, we do not exclude the possibility of \textquotedblleft puddles" of uniform charge in multilayer graphene systems due to the presence of additional Hartree terms in the stacking direction (see for example Ref.~\cite{Lewandowsky2023Electrostatic}) where the electrostatic interactions may be strong enough to induce large Umklapp processes that may give rise to a larger superconducting critical temperature.

\section{Evolution of the density of states with perpendicular electric field}
\label{sec:ElectricField}
\begin{figure}[ht!]
\centering
\includegraphics[scale = 0.25]{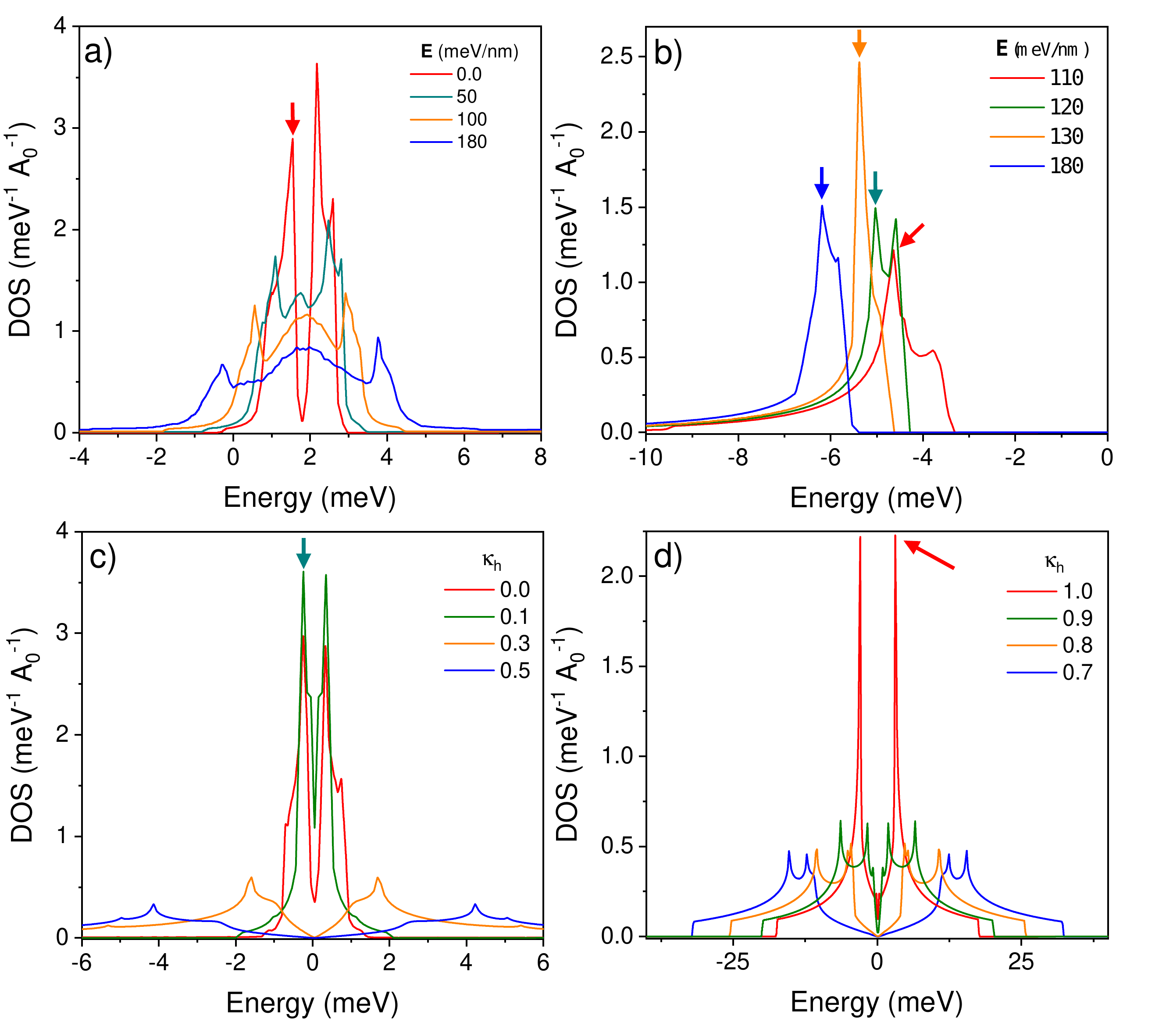}
\caption{Density of states of a) TBG and b) TDBG as a function of the external electric field. In c) and d) we display the DOS of TTG as a function of the coupling parameter for $\theta=1.1^\circ$ and $\theta=1.9^\circ$, respectively. Colored arrows indicates the vHS chosen to calculate the critical temperature.}
\label{fig: DosBias}
\end{figure}

In Fig.~\ref{fig: DosBias}, we illustrate the evolution of the density of states as a function of the coupling parameter and electric field a) TBG and b) TDBG, c) hTTG with $\theta=1.1^\circ$ and d) hTTG with $\theta=1.9^\circ$. In TBG, the external potential couples with different signs to each Dirac cone. Due to the layer inversion symmetry, this coupling does not create a gap. Instead, the Dirac cones shift in opposite directions, leading to a reduction and spreading of the DOS peaks.~\cite{Koshino2014Optical}. In the case of TDBG, as shown in Fig.~\ref{fig: DosBias}b), the DOS is sensitive to the electric field. At the maximum value considered, $\mathbf{E} = 180$ meV/nm, the DOS exhibits significant reduction due to an strong overlap between both middle bands. As shown in Fig.~\ref{fig: DosBias}c), in the transition from TBG to hTTG with $\theta=1.1^\circ$, the DOS is strongly suppressed with the coupling because an increasing in the bandwidth. For $\kappa=0.3$, the DOS reduction is several times smaller than that of the TBG limit. In addition, for the transition from TBG to hTTG with $\theta=1.90^\circ$, the behaviour is the opposite. In the TBG limit, the bandwidth of the narrow bands is $\sim 200$ meV. As the coupling increases, Fig.~\ref{fig: DosBias}d), the DOS is enhanced due to a band flattening induced by the hybridization with the additional Dirac cone. In the hTTG limit, the bandwidth is reduced to $\sim 30$ meV and the DOS has large peaks.  

\section{Evolution of the band structure with the coupling parameter}
\label{sec:HartreeEvolution}
\subsection{Evolution from TBG to TDBG}

\begin{figure*}[ht!]
    \centering
    \includegraphics[width = \textwidth]{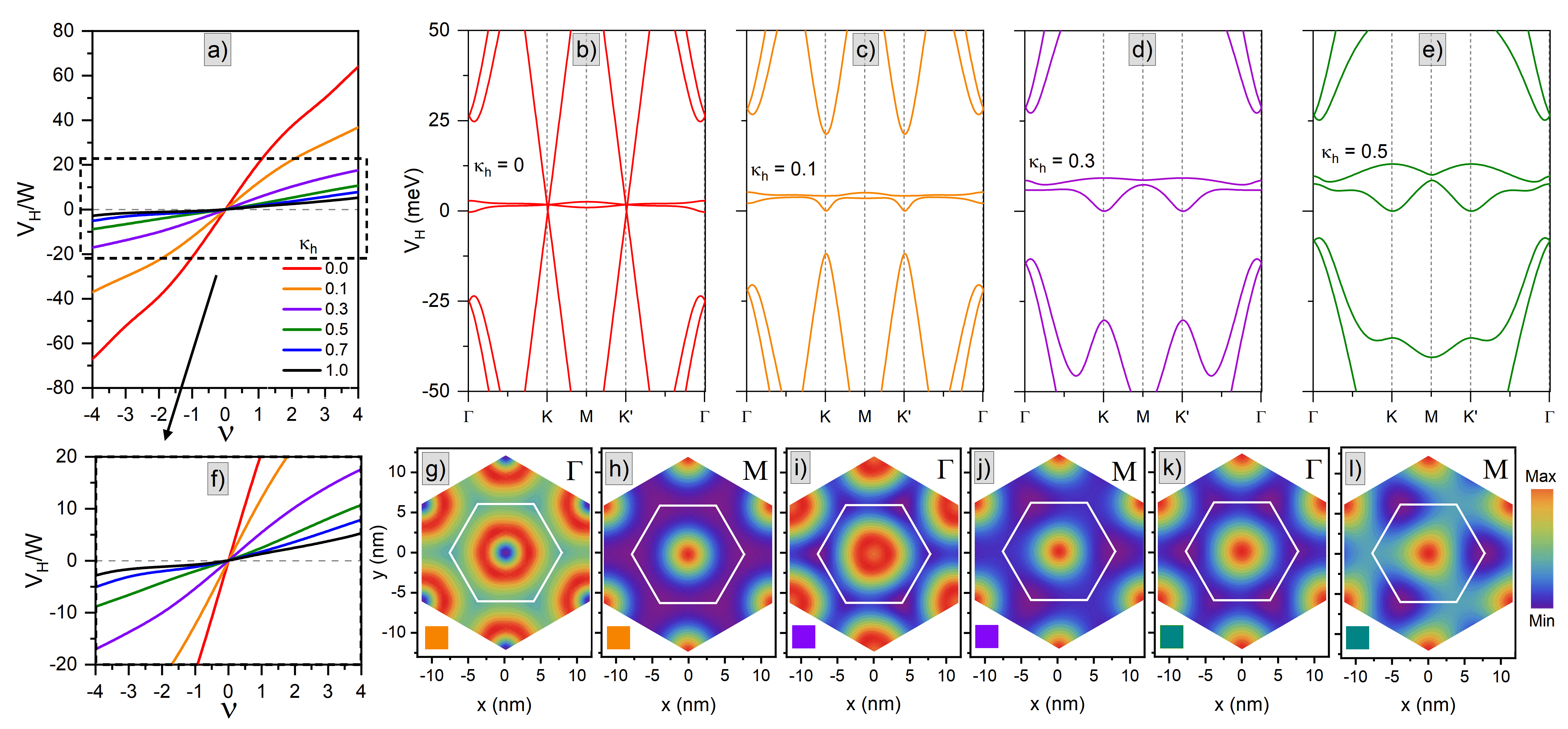}
    \caption{Band structure and charge density of TDBG with $\mathbf{E}=0$, as a function of the coupling scaling parameter $\kappa_h$. a) Hartree strength versus filling fraction $\nu$, where $\nu=-4 (4)$ for fully empty (filled) middle bands. An enlarged region (dashed square) is shown in f). Panels b) to e) display the evolution of the band structure at charge neutrality as a function of  $\kappa_h$. Panels g) to l) are the charge densities in real space at the points $\Gamma$ and $M$ for each value of $\kappa_h$ indicated by the colored square which follows the color scheme in a) and b)-e).}   
    \label{fig:HartreeTBG}
\end{figure*}

Figure~\ref{fig:HartreeTBG} illustrates the band structure, charge density, and Hartree potential for various values of the coupling scaling parameter. In the TBG limit ($\kappa_h=0$), the band structure results from the diagonalization of a decoupled system comprising two graphene monolayers and a twisted bilayer graphene. The Hartree strength is given by its maximum value $V_H(0,0)\equiv V_H$ in Eq.~\ref{eq: VhRealSpace} normalized by the bandwidth, this is $V_H/W$. Here, $W$ is the bandwidth of the non interacting bands. In Fig.~\ref{fig:HartreeTBG}a) we display the Hartree strength. The bandwidth for each coupling strength is calculated by considering both narrow bands. As the coupling strength increases, the Hartree strength is strongly suppressed due to a charge redistribution to the outer layers. The evolution of the band structure is depicted in (b)-(e), along with the corresponding charge distribution at $\Gamma$ and $M$ in (g)-(l).    
 
\subsection{Evolution from TBG to hTTG}

\begin{figure*}[ht!]
    \centering
    \includegraphics[width = \textwidth]{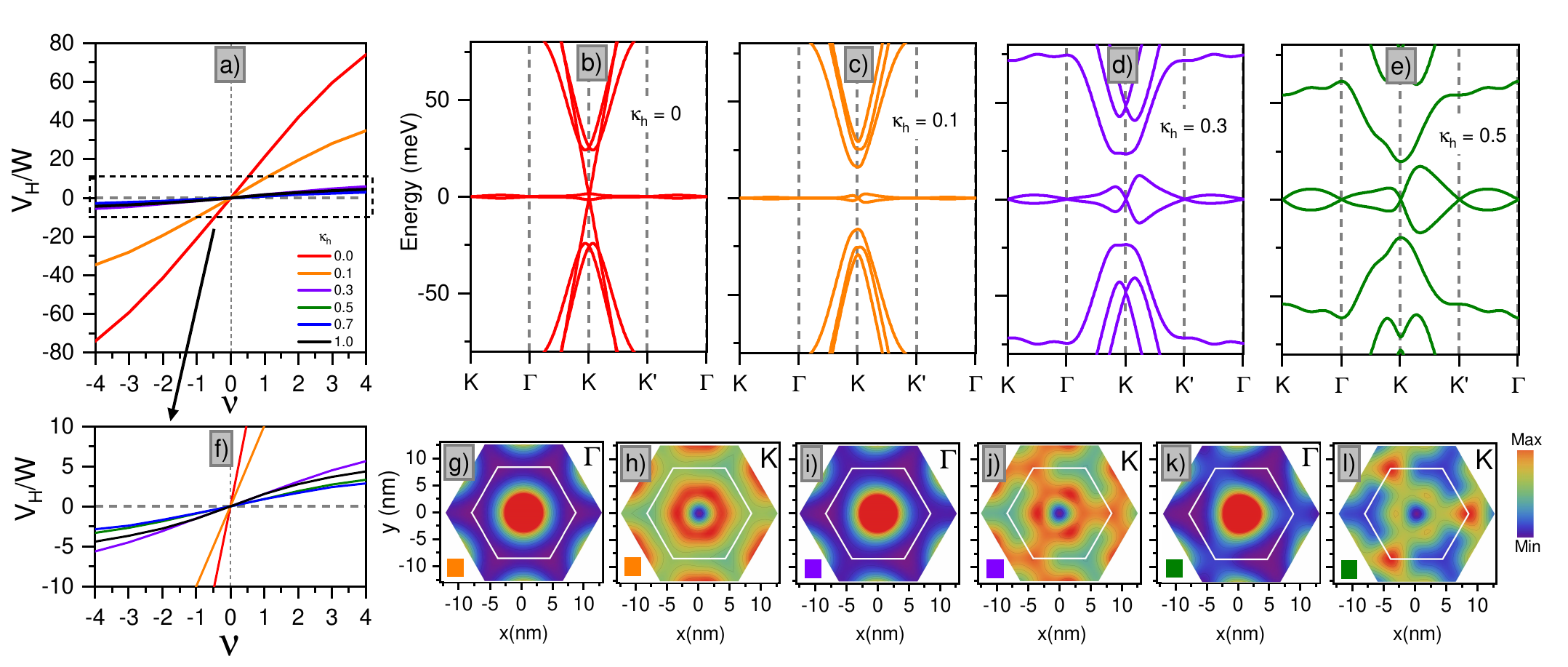}
    \caption{Band structure and charge density of hTTG as a function of the coupling scaling parameter $\kappa_h$ for a twist angle $\theta = 1.10^\circ$. a) Hartree strength versus filling fraction $\nu$, where $\nu=-4 (4)$ for fully empty (filled) middle bands. An enlarged region (dashed square) is shown in f). Panels b) to e) display the evolution of the band structure at charge neutrality as a function of  $\kappa_h$. Panels g) to l) are the charge densities in real space at the points $\Gamma$ and $K$ for each value of $\kappa_h$ indicated by the colored square which follows the color scheme in a) and b)-e).}   
    \label{fig:HartreeTTG1}
\end{figure*}

\begin{figure*}[ht!]
    \centering
    \includegraphics[width = \textwidth]{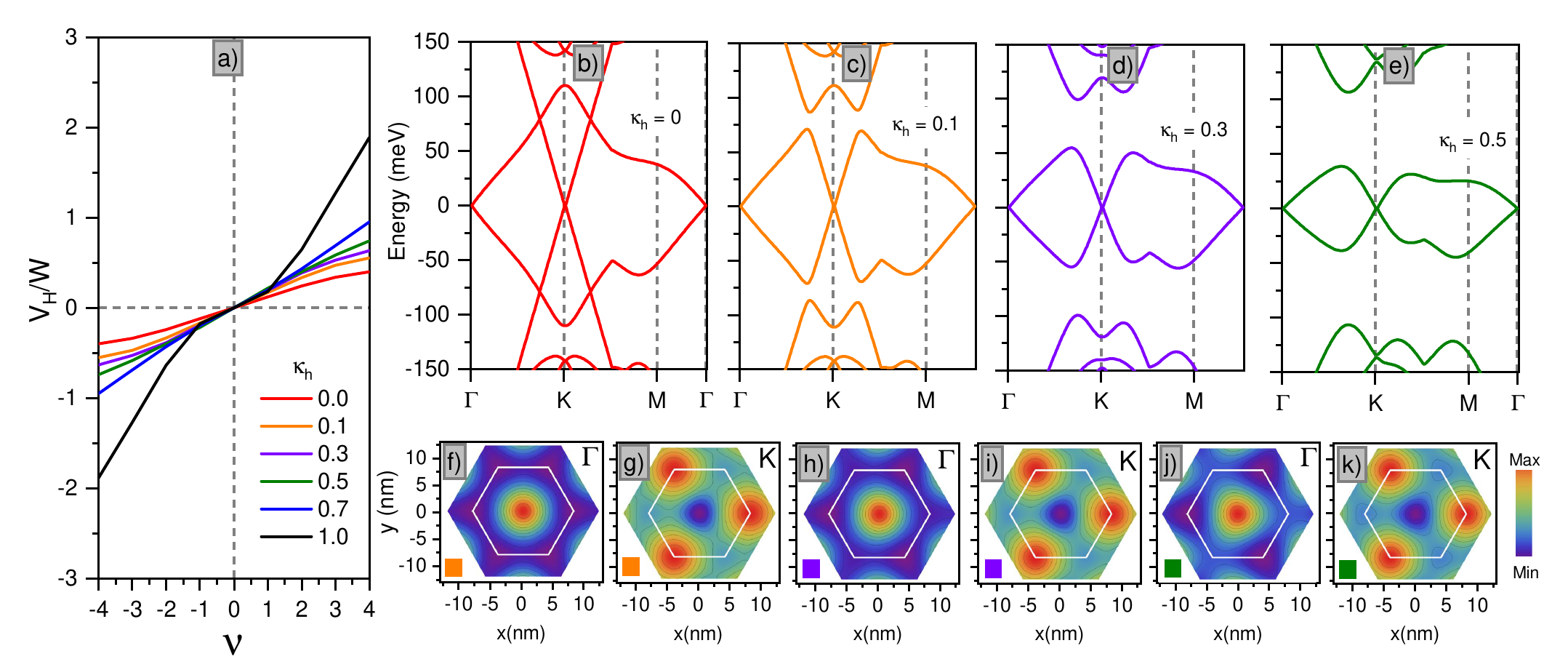}
    \caption{Band structure and charge density of hTTG as a function of the coupling scaling parameter $\kappa_h$ for a twist angle $\theta = 1.90^\circ$. a) Hartree strength versus filling fraction $\nu$, where $\nu=-4 (4)$ for fully empty (filled) middle bands. Panels b) to e) display the evolution of the band structure at charge neutrality as a function of  $\kappa_h$. Panels f) to k) are the charge densities in real space at the points $\Gamma$ and $K$ for each value of $\kappa_h$ indicated by the colored square which follows the color scheme in a) and b)-e).}   
    \label{fig:HartreeTTG2}
\end{figure*}

We discuss the effect of a transition from TBG to hTTG. We consider two situations: the first one is a transition from TBG magic angle $\theta=1.1^\circ$ to hTTG, as shown in Fig.~\ref{fig:HartreeTTG1}. The second one is a transition from TBG to {\it magic angle} hTTG~\cite{Foo2023TTG,Yang2023Multi}, which for our set of parameters is at $\theta=1.90^\circ$, as shown in Fig.~\ref{fig:HartreeTTG2}. In the first case, fig.~\ref{fig:HartreeTTG1} illustrates the evolution of the Hartree potential as a function of the coupling scaling parameter. In the limiting case $\kappa_h=0$, the band structure results from the diagonalization of a decoupled system consisting of a graphene monolayer and a magic angle twisted bilayer graphene (TBG). The Hartree potential strength is strongly reduced by increasing coupling strength. For $\kappa_h > 0.1$ the Hartree strength is negligible and the twisted system is no longer a superconductor, see Fig.~\ref{fig:SChTTG110}. For larger values of coupling we did not found any additional SC state. Our results suggest that the strong charge redistribution and the spreading of the DOS, Fig.~\ref{fig: DosBias}c), with coupling in the trilayer system strongly suppresses the KL superconductivity.  

\section{Evolution of superconductivity from TBG to \texorpdfstring{\lowercase{h}{T}} TTG at \texorpdfstring{$\theta = 1.10^\circ$}{}}
\begin{figure*}[ht!]
    \centering
    \includegraphics[scale=0.6]{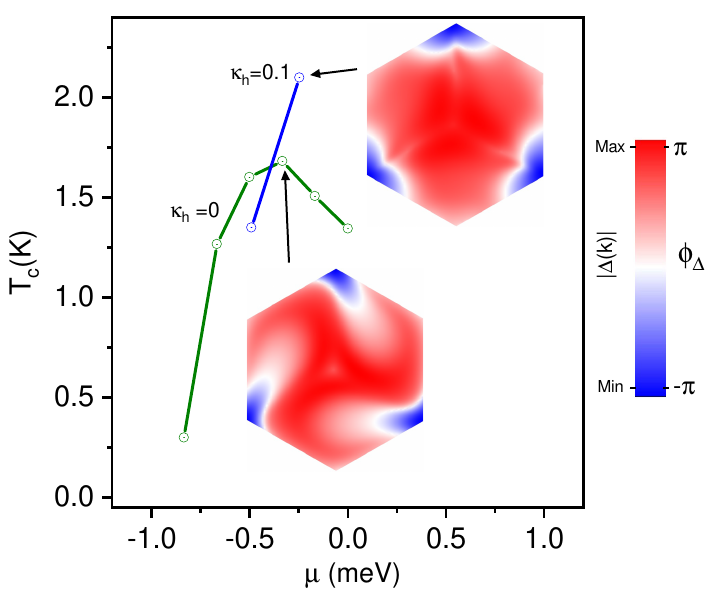}
    \caption{Evolution of superconductivity from TBG to hTTG at $\theta = 1.10^\circ$. Green line is the SC dome for TBG+2MG, $\kappa=0$ and blue line for $\kappa=0.1$ where the SC increases due to the small coupling. We cannot find any SC state on another points. For $\kappa>1$ no SC is found. Note that we find SC in hTTG at $\theta = 1.10^\circ$ as described in the main text.} \label{fig:SChTTG110}
\end{figure*}

In Fig.~\ref{fig:SChTTG110} we show the evolution of the critical superconducting dome as a function of the coupling parameter for a transition from TBG to hTTG. The behaviour is similar to the case of small $\kappa$ in Fig. \textcolor{red}{3}, except that we cannot find superconductivity for $\kappa>0.1$. This behaviour can be compare with Fig.~\ref{fig: DosBias}c) and Fig.~\ref{fig: DelocalizationTTGa} where the DOS is broadened and the MV parameter increases (see also Fig.\textcolor{red}{4}a) in the main text).   

\begin{figure*}[t!]
    \centering
    \includegraphics[width = \textwidth]{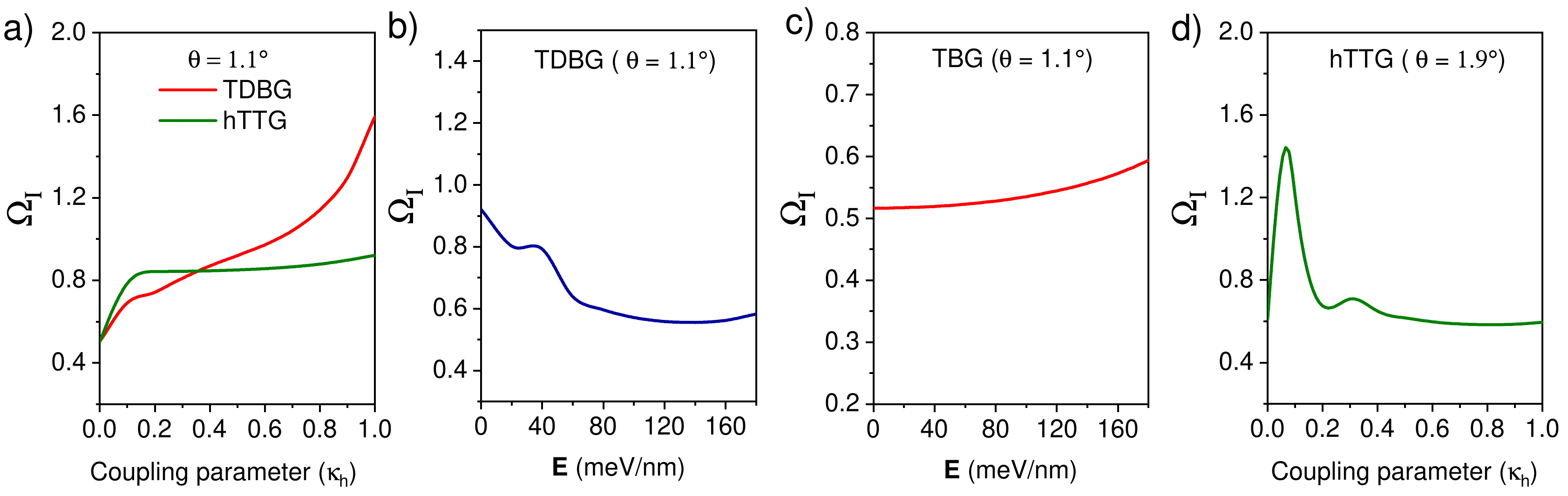}
    \caption{Evolution of the Marzari-Vanderbilt cumulant $\Omega_{\mathrm{I}}$. a) Transition from TBG to TDBG and hTTG for $\theta=1.1^\circ$ and $\vb{E}=0$. b) and c) display the variation of the cumulant as a function of the electric field in TDBG and TBG, respectively. d) Transition from TBG to hTTG for $\theta=1.9^\circ$ and $\vb{E}=0$. 
    } 
    \label{fig: cumulant}
\end{figure*}

\section{Wavefunction homogeneity}
\label{sec:ChargeLocalization}

\subsection{Marzari-Vanderbilt cumulant}
While the Hartree potential give us an insight of the overlaps between Umklapp processes, a useful quantity for defining the degree of homogeneity of the wavefunctions is the Marzari-Vanderbilt cumulant~\cite{Marzari1997Maximally}. By expressing
the metric tensor as $g_{\mu\nu}=\mathrm{Re}\left\langle \partial_{\mu}u_{\boldsymbol{k}}\left|Q_{\boldsymbol{k}}\right|\partial_{\nu}u_{\boldsymbol{k}}\right\rangle $
where $Q_{\boldsymbol{k}}=1-\left|u_{\boldsymbol{k}}\right\rangle \left\langle u_{\boldsymbol{k}}\right|$
and $\partial_{\mu}=\partial/\partial k_{\mu}$, the Marzari-Vanderbilt
cumulant is given by 
\begin{equation}
\Omega_{\mathrm{I}}=\frac{A_{0}}{(2\pi)^{2}}\intop_{BZ}d\boldsymbol{k}\mathrm{Tr}\left[g(\boldsymbol{k})\right]=\frac{1}{N_{c}}\sum_{\boldsymbol{k}}\mathrm{Tr}\left[g(\boldsymbol{k})\right],
\label{eq: Cumulant1}
\end{equation} 
 where $\mathrm{Tr}\left[g\right]=g_{xx}+g_{yy}$ is the trace of
the metric tensor and $N_{c}$ is the number of k-points in the numerical
grid~\cite{Marzari1997Maximally}.  By calculating $\Omega_{\mathrm{I}}$ as a function of $\kappa_h$, we can analyze how the homogeneity of states changes. This, in conjunction with the Hartree calculations, allows us to estimate the presence of a superconducting behavior. Let's start with the physical meaning of this factor: the Marzari-Vanderbilt cumulant is the sum of the trace of the metric tensor over BZ. In calculating metrix tensor, we take the difference between one wavefunction (says $| u_{\boldsymbol{k}}\rangle$) and it's neighbors, through the projection operator $Q_{\boldsymbol{k}}=1-\left|u_{\boldsymbol{k}}\right\rangle \left\langle u_{\boldsymbol{k}}\right|$, this difference is projected into the orthogonal subspace of $| u_{\boldsymbol{k}}\rangle$. Then the sum of the trace of the metric tensor measures the intrinsic {\it smoothness} of the underlying Hilbert space and the trace of the metric tensor at each value of $\boldsymbol{k}$ measures the degree of
mismatch between the neighboring Bloch subspaces $\boldsymbol{k}$ and $\boldsymbol{k+q}$. The MV parameter measures the {\it similarity} of the states on one or in composite bands.  In addition, both in susceptibility and gap equation, the overlap of states $\langle u_{n,\mathbf{k}}|u_{m,\mathbf{k+q}}\rangle $ weights the coupling process, and therefore this overlap measures the similarity of $|u_{m,\mathbf{k+q}}\rangle$ and $|u_{n,\mathbf{k}}\rangle$, which is also captured by MV factor.

\subsection{TBG and TDBG}

\begin{figure}[ht!]
    \centering
    \includegraphics[scale = 0.40]{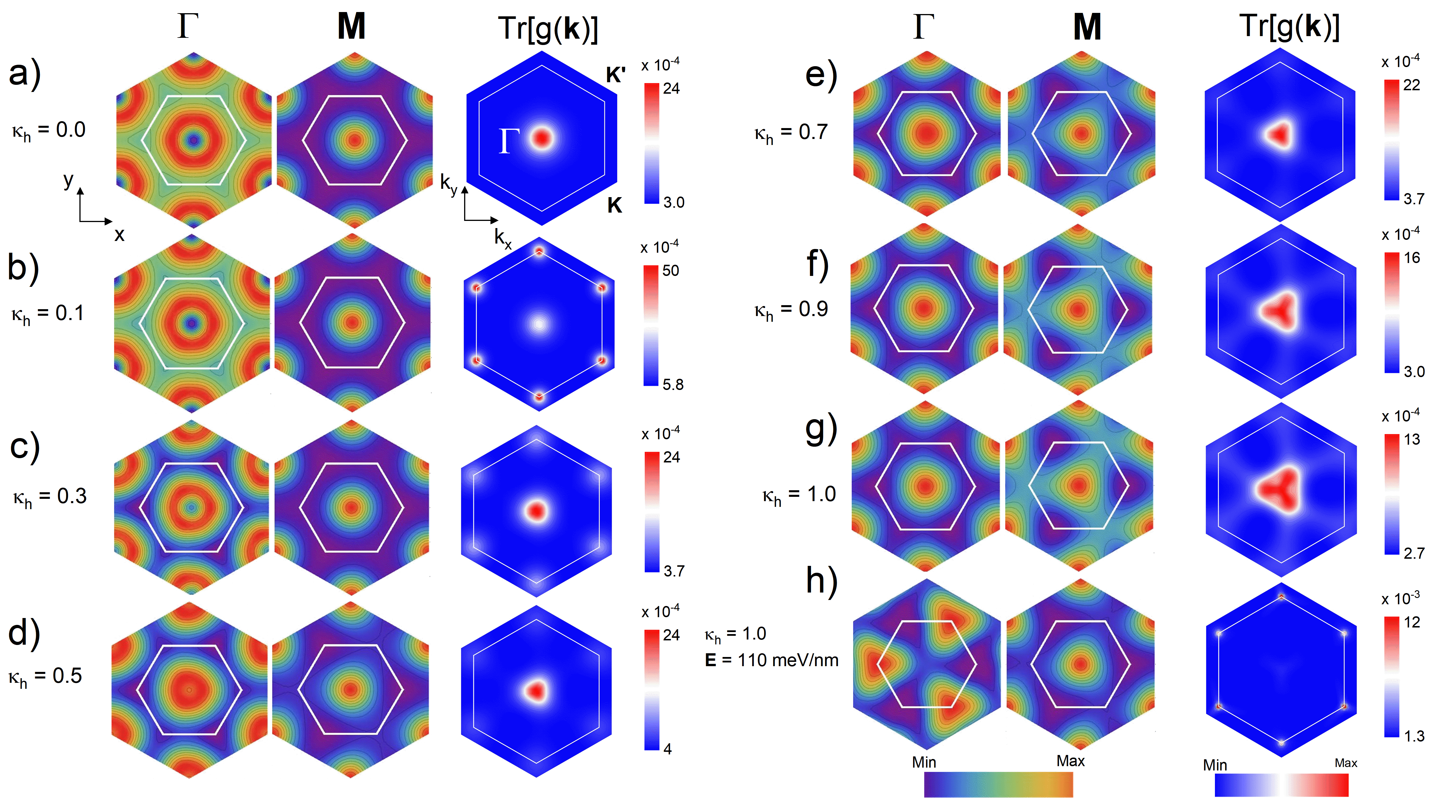}
    \caption{Charge distribution as a function of the coupling parameter in TDBG. First two columns are the charge density of states in real space, the last column is $\mathrm{Tr}\left[g(\boldsymbol{k})\right]$ in the mBZ. 
    The variation of the charge localization in real space is calculated at $\Gamma$ and $M$. In the distribution of the cumulant, $\mathrm{Tr}\left[g(\boldsymbol{k})\right]$, red regions are those where the wavefunctions are different from their neighbours. In h) we display the case of $\mathbf{E} = 110$ meV/nm and $\kappa_h = 1$, the cases of $\mathbf{E} = 120, 130 \text{ and } 180$meV/nm display a very similar behavior. Rows a) and g) are the limiting cases for TBG and TDBG, respectively.} 
    \label{fig: DelocalizationTDBG}
\end{figure}

\begin{figure}[ht!]
    \centering
    \includegraphics[scale = 0.40]{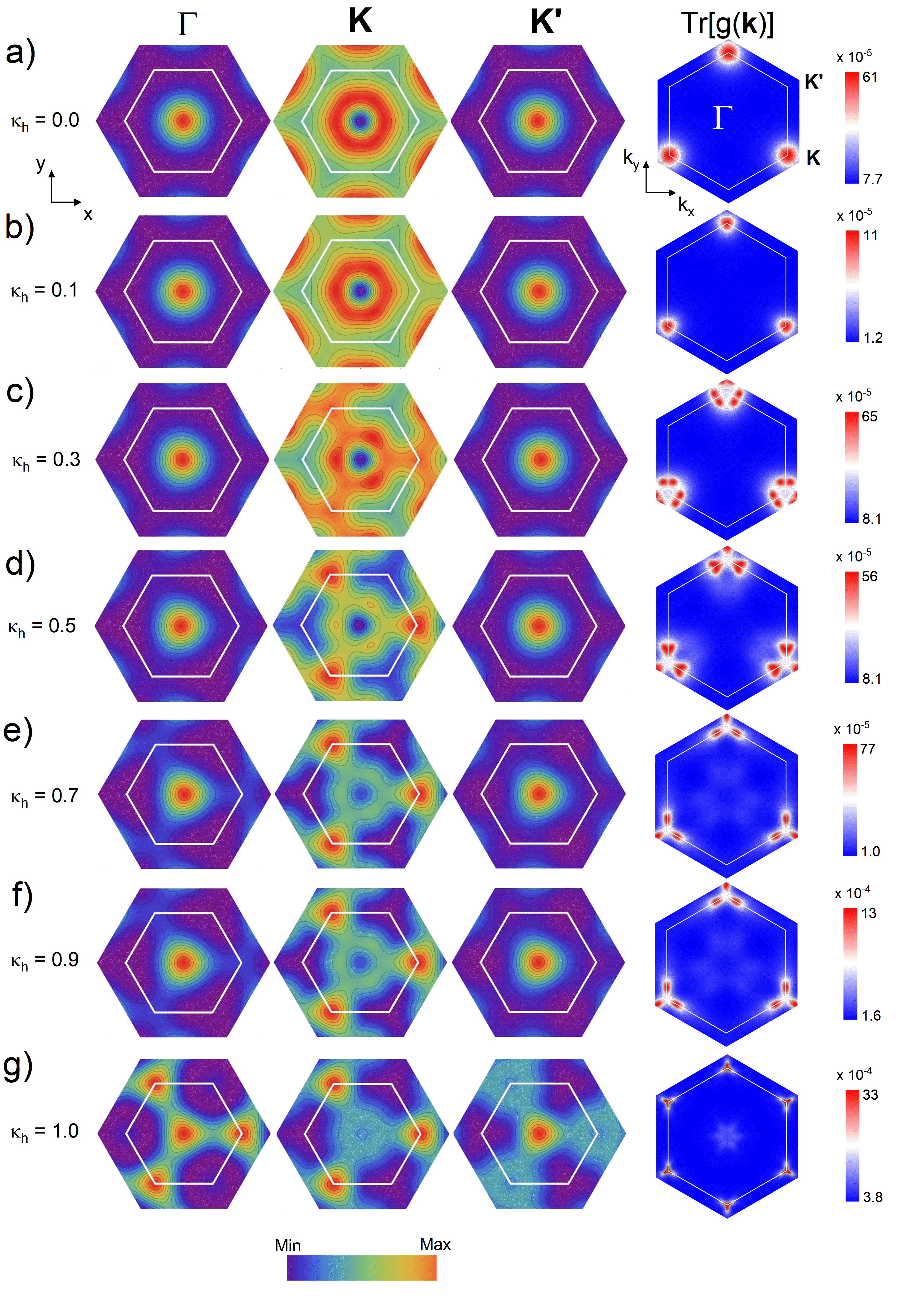}
    \caption{Charge distribution as a function of the coupling parameter in hTTG for $\theta=1.1^\circ$. The first three columns are the charge density of states in real space, and the last column is $\mathrm{Tr}\left[g(\boldsymbol{k})\right]$ in the mBZ.  
    The variation of the charge localization in real space is calculated at $\Gamma$ and $K$ and $K'$. In the distribution of the cumulant, $\mathrm{Tr}\left[g(\boldsymbol{k})\right]$, red regions are those where the wavefunctions are different from their neighbours. Rows a) and g) are the limiting cases for TBG and hTTG, respectively.} 
    \label{fig: DelocalizationTTG}
\end{figure}

In our systems, the parameter $\Omega_{\mathrm{I}}$ varies with the coupling parameter $\kappa_{h}$ as it quantifies wavefunction homogeneity. Thus, we can compare its variation with $\kappa_{h}$ as we transform one system to another. Figure~\ref{fig: cumulant}a) and c) shows the variation of $\Omega_{\mathrm{I}}$ as a function of the coupling parameter, and Fig.~\ref{fig: cumulant}b) and d) as a function of the electric field. In the totally decoupled case $\kappa_{h}=0$, we recover the TBG value, here, even though the wavefunctions are strongly localized at the AA centers for values of momenta at the edges of the mBZ, the wavefunctions are more delocalized around $\Gamma$ due to an  hibridization with the remote bands~\cite{Po2019Faithful}. This is the reason of the maxima of Tr$[g(\boldsymbol{k})]$ at $\Gamma$ in Fig.~\ref{fig: DelocalizationTDBG}a) and $K$ Fig.~\ref{fig: DelocalizationTTG}a). We note that in the decoupled situations, both TDBG and hTTG are equivalent up to a shift of the corresponding mBZ, so that the $\Gamma$ point for TDBG is equivalent to the $K$ point in hTTG. In the case of TDBG, we found SC after introducing an external electric field (see main text). In this situation, as shown in Fig.~\ref{fig: cumulant}b), the charge becomes more localized as a function of the field. In contrast, in TBG, an external electric field produces dispersive narrow bands~\cite{Koshino2014Optical}, as shown in Fig.~\ref{fig: cumulant}c). 

\subsection{hTTG}

\begin{figure}[ht!]
    \centering
    \includegraphics[scale = 0.40]{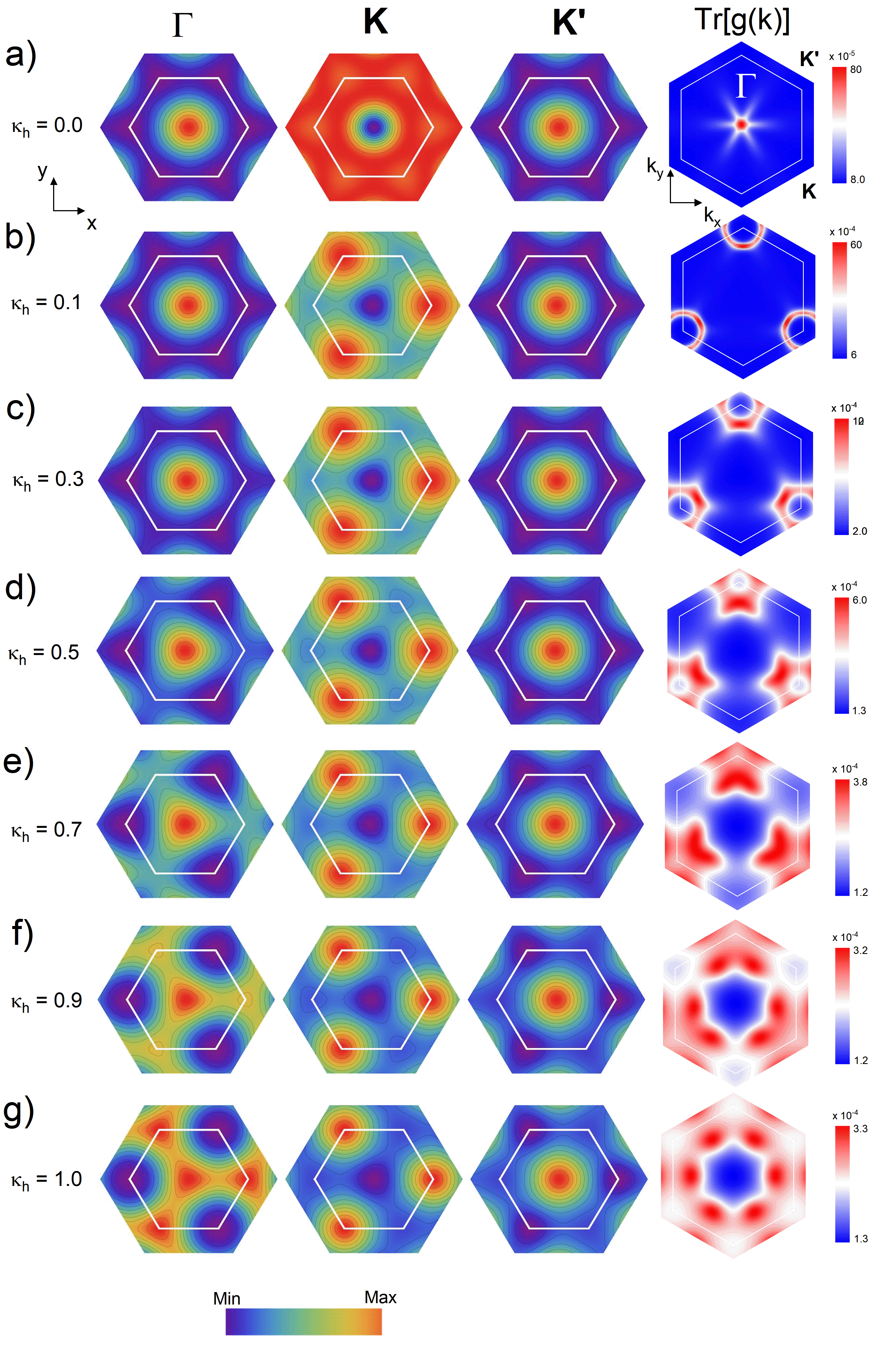}
    \caption{Charge distribution as a function of the coupling parameter in hTTG for for $\theta=1.9^\circ$. First three columns are the charge density of states in real space, the last column is $\mathrm{Tr}\left[g(\boldsymbol{k})\right]$ in the mBZ.  
    The variation of the charge localization in real space is calculated at $\Gamma$ and $K$ and $K'$. In the distribution of the cumulant, $\mathrm{Tr}\left[g(\boldsymbol{k})\right]$, red regions are those where the wavefunctions are different from their neighbours.} 
    \label{fig: DelocalizationTTGa}
\end{figure}

In the transition from TBG to hTTG, wavefunction localization depends on the twist angle. As described in Ref.~\cite{Foo2023TTG}, the hTTG magic angle is different from the usual $\theta=1.1^\circ$ TBG case. In our parameterization, the magic angle is at $\theta=1.9^\circ$. We describe two different situations that we believe also summarize intermediate situations: Firstly, we start from the TBG magic angle, where there is SC, strong Hartree interactions, and a small MV parameter. By increasing the coupling $\kappa_h$, both SC and Hartree interactions are suppressed. As shown in Fig.~\ref{fig:SChTTG110}, the superconducting (SC) dome for SC TBG is clear, but for larger values, no SC is found. The MV parameter, Fig.~\ref{fig: cumulant} rapidly increases, consistent with charge delocalization behavior. Secondly, we start from a TBG at a large angle, $\theta=1.9^\circ$, with no Hartree effects nor SC behavior and also with strong charge delocalization. The variation of $\kappa_h$ is not trivial since it involves overlaps between the wide bands and the Dirac cones, as shown in Fig.~\ref{fig:HartreeTTG2}. In the limit of $\kappa_h = 1$, the MV parameter is still small but uniformly distributed within the mBZ, see scale in Fig.~\ref{fig: DelocalizationTTGa}g), the Hartree strength, even though it is small, it becomes comparable with the bandwidth of the narrow bands due to an enhancement of the density of states (DOS), as shown in Fig.~\ref{fig: DosBias}. Therefore, non-SC TBG can be converted into a SC hTTG with uniform quantum metric. The latter suggest that magic angle hTTG is a good candidate for a fractional Chern insulating behaviour~\cite{Devakul2023Magic,Ledwith2020Fractional}. 

\section{Screening and linearized gap equation}
\label{sec:Screening}
Here we provide the details concerning the calculation of the screened Coulomb potential in the momentum space and the linearized gap equation used to compute the critical temperature and order parameter of the superconducting instability. We assume that the bare Coulomb potential at momentum $\vb{q}$ takes the form,
\begin{equation}
    \label{eq:bare_coulomb_pot}
    \mathcal{V}_{C}(\vb{q}) = \frac{2\pi e^{2}}{\epsilon} \frac{\tanh{(d_{g}|\vb{q}|})}{|\vb{q}|},
\end{equation}
where $e$ is the electron charge, $d_g$ is the distance from the sample to the metallic gates and $\epsilon$ is the relative dielectric constant of the environment. In this work we use $\epsilon = 10$ and $d_g = 40$ nm. \\

We adopt the RPA to compute the screened Coulomb potential which ultimately enables the formation of Cooper pairs. We first consider the effect of pure electronic correlations and compute the bare electronic susceptibility, which is the density-density response function due to the electron-hole excitations or bubble diagrams \cite{cea21Coulomb,cea2022superconductivity,JimenoPozo2023,ZiyanLi2023}.

The order of discussion below is as follows: first, we review the charge density and susceptibility in a normal crystal; then, we transition to the moiré system. Finally, we emphasize that the susceptibility in the moiré system is a generalization of that in a normal crystal. The susceptibility in normal (non moiré) crystals is given by~\cite{cea2022superconductivity}:

\begin{equation}
\begin{split}
    \chi_0(\mathbf{q}) =& \frac{4}{V_{mBZ}}\sum_{k,m,n}\frac{f(\epsilon_{n,\mathbf{k}})-f(\epsilon_{m,\mathbf{k+q}})}{\epsilon_{n,\mathbf{k}}-\epsilon_{m,\mathbf{k+q}}} 
    \langle u_{n,\mathbf{k}}|u_{m,\mathbf{k+q}}\rangle 
    \langle u_{m,\mathbf{k+q}}|u_{n,\mathbf{k}}\rangle , 
\end{split}
\end{equation}
where $V_{mBZ}$ is the area of the Brillouin zone (BZ), and $u_{n,\mathbf{k}}$ are the eigenstates of the Hamiltonian labelled by momentum $\mathbf{k}$ and band index $n$. By recalling that the charge density has the form $\delta\rho(\vb{q}) = \sum_{\vb{q}} \langle u_{m,\mathbf{k+q}}|u_{n,\mathbf{k}}\rangle $ 
with $\mathbf{k}$ running over BZ, and by recognizing the overlap $\langle u_{m,\mathbf{k_0+q}}|u_{n,\mathbf{k_0}}\rangle$ as an individual contribution from momentum $\mathbf{k_0}$ to the charge density, it is then clear that the susceptibility is given by the product of such individual contributions weighted by a Lindhard factor $[f(\epsilon_{n,\mathbf{k}})-f(\epsilon_{m,\mathbf{k+q}})]/(\epsilon_{n,\mathbf{k}}-\epsilon_{m,\mathbf{k+q}}) $. This justifies the name density-density response function for $\chi_0$.
In a moir\'e system, as shown before, the charge density depends on the momentum $\mathbf{q}$ as well as on the reciprocal lattice vectors:

\begin{equation}
    \delta\rho(\vb{q}+\vb{G}) = \frac{1}{V_{mBZ}}\sum_{m,n,\alpha}\int_{\text{mBZ}}\dd{\vb{k}}\langle u_{m}^{\alpha}(\vb{k}+\vb{q})|M(\vb{G})|u_{n}^{\alpha}(\vb{k})\rangle,
\end{equation}

The density-density response process in a moir\'e system is more complex compared to a non-twisted system. Since the Umklapp processes allow interactions with momentum differences of a reciprocal lattice vector and the charge density depends on $\mathbf{G}$, this makes the Umklapp processes non-trivial. The electronic susceptibility $\chi_0(\mathbf{q})$ accounts for the hopping processes from states with momentum $\mathbf{k}$ to $\mathbf{k+q}$, with k running over all the BZ. Considering one contribution of the process from $\mathbf{k_0}$ to $\mathbf{k_0+q}$, the Umklapp process allows interaction up to a reciprocal lattice vector difference, thus the hopping from $\mathbf{k_0+G}$ to $\mathbf{k_0+q + G'}$ should be taken into account. Remembering that the bare electronic susceptibility is the product of the charge density weighted by the Lindhard factor and noting that we already introduced the connection matrix to express the charge density, the component of the susceptibility that corresponds to the process we consider is then given by

\begin{align}
\left[\chi_{0}(\boldsymbol{q})\right]_{\boldsymbol{G_i,}\boldsymbol{G_j}}=\frac{4}{V_{mBZ}}\sum_{\boldsymbol{k},m,n}\frac{f(\epsilon_{n,\boldsymbol{k}})-f(\epsilon_{m,\boldsymbol{k}+\boldsymbol{q}})}{\epsilon_{n,\boldsymbol{k}}-\epsilon_{m,\boldsymbol{k}+\boldsymbol{q}}}\left\langle u_{n,\boldsymbol{k}}\left|M^\dagger(\boldsymbol{G}_{i})\right|u_{m,\boldsymbol{k+q}}\right\rangle \left\langle u_{m,\boldsymbol{k+q}}\left|M(\boldsymbol{G}_{j})\right|u_{n,\boldsymbol{k}}\right\rangle,
\label{eq:SusGGp}
\end{align}
where we introduce a factor of 2 to take into account degeneracy of spin and valley. From the above expression, it is clear that the susceptibility is promoted to a matrix because of the Umklapp process, in other words, the susceptibility now is a rank-2 tensor field over the mBZ. The connection matrix allows us to write the bare electronic susceptibility matrix as a tensor product,
\begin{equation}
\begin{split}
    \mathcal{X}_0(\mathbf{q}) =& \frac{4}{V_{mBZ}}\sum_{k,m,n}\frac{f(\epsilon_{n,\mathbf{k}})-f(\epsilon_{m,\mathbf{k+q}})}{\epsilon_{n,\mathbf{k}}-\epsilon_{m,\mathbf{k+q}}} 
    \langle \Psi_{n,\mathbf{k}}||\Psi_{m,\mathbf{k+q}}\rangle \otimes
    \langle \Psi_{n,\mathbf{k}}||\Psi_{m,\mathbf{k+q}}\rangle^\dagger , 
\end{split}
\label{eq:SusceptibilityMoire}
\end{equation}
where $f$ refers to the Fermi-Dirac distribution and $\epsilon_{n,\vb{k}} = E_{n,\vb{k}} - \mu$, where $E_{n,\vb{k}}$ is the energy and $\mu$ the Fermi energy. We introduce an overlap vector to account for the Umklapp processes, given by,
\begin{equation}
 \langle \Psi_{n,\mathbf{k}}||\Psi_{m,\mathbf{k+q}}\rangle = \begin{pmatrix}
                       \langle u_{n,\mathbf{k}}|M^\dagger(\vb{G_1})|u_{m,\mathbf{k+q}}\rangle \\
                        \langle u_{n,\mathbf{k}}|M^\dagger(\vb{G_2})|u_{m,\mathbf{k+q}}\rangle \\
                        \vdots \\
                        \langle u_{n,\mathbf{k}}|M^\dagger(\vb{G_{n}})|u_{m,\mathbf{k+q}}\rangle
                                                            \end{pmatrix}.
\label{eq: matrix element chi}
\end{equation}
We have found that the overlap $\langle u_{m,\mathbf{k+q}}|M(G_i)|u_{n,\mathbf{k}}\rangle$ becomes exponentially small as we connect additional $\vb{G}$ couplings. In practice, we cut off the couplings within the second nearest reciprocal lattice vectors. By restricting the number of reciprocal lattice vectors to one, meaning we only consider $\mathbf{G} = \mathbf{0}$, the corresponding connection matrix is the identity matrix. This brings us back to the normal crystal case. The susceptibility in Eq.~\ref{eq:SusceptibilityMoire} for moiré systems then can be viewed as a generalization of that of a normal crystal.

\begin{figure}[ht!]
    \centering
    \includegraphics[width = 0.40\textwidth]{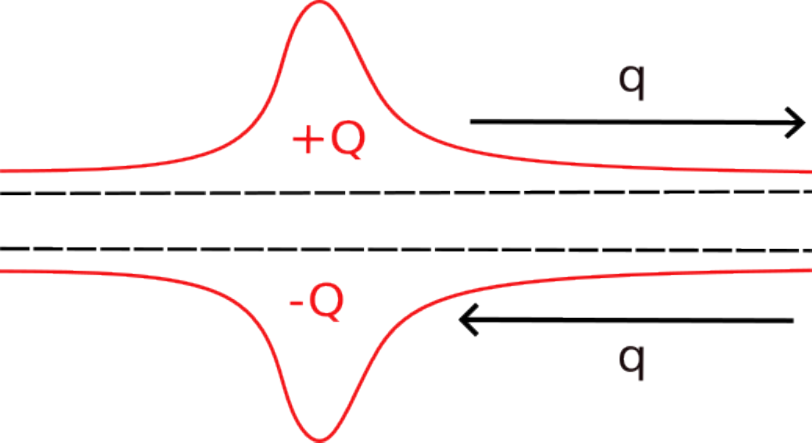}
    \caption{Illustration of odd combination of phonons, which results in accumulation of opposite sign of charges in different layer. Thus only generates a local potential}
    \label{fig: oddcombination}
\end{figure}

We then consider the effect of the phonon-mediated electron-electron interactions. In the following we neglect pairing due to the bare electron-phonon coupling because it is weak compared with electron-electron interaction mediated by acoustic phonons~\cite{cea21Coulomb}. Instead, we consider that longitudinal acoustic phonons couple to electrons through the local compression and expansion that they induce. We then compute how this phonon-mediated electron-electron interaction affects the bare electronic susceptibility. In the RPA limit the matrix elements of the screened electronic susceptibility is given by,

\begin{equation}
    \chi_{\text{ph}}(\vb{q}) = \chi_{0}(\vb{q})\left(I - \mathcal{V}_{ph}(\vb{q},\omega)\chi_{0}(\vb{q})\right)^{-1}
\end{equation}
where $I$ is a identity matrix of suitable dimensions and $\mathcal{V}_{ph}(\vb{q},\omega)$ is the potential of the interaction, we note that $\chi_0(\vb{q})$ is a matrix and we promote bare Coulomb potential to a matrix with the form $\mathcal{V}_{C}(\vb{q})_{\vb{G_1},\vb{G_2}} = \delta_{\vb{G_1},\vb{G_2}}\mathcal{V}_{C}(\vb{q}+\vb{G_1})$, where the expression of the Coulomb potential is given in the beginning. Then, the multiplication and inversion in these equations is an operation on a matrix,
\begin{equation}
    \label{eq:electron_phonon_pot}
    \mathcal{V}_{ph}(\vb{q},\omega) = \frac{D_0^2|\vb{q}|^2}{\rho(\omega^2-\omega^2_q)},
\end{equation}
where $\rho$ is the mass density, $\omega_q = v_s|\vb{q}|$ is the frequency of a phonon with momentum $\vb{q}$, and $v_s = \sqrt{(\lambda+2\mu)/\rho}$ is the velocity of sound, $\lambda$ and $\mu$ being the elastic Lam\'e coefficients. We neglect the coupling between acoustic phonons in the two layers. This approximation is valid when the atomic positions in the layers are not significantly modified by relaxation effects, which happens when the twist angle is small. Electrons, delocalized through the two layers, can couple to the even and the odd combination of phonons in each layer. The odd combination leads to a change in the chemical potential of the opposite sign in the two layers. As a result, the total induced charge vanishes and this phonon superposition does not lead to a long-range electrostatic potential, as shown in Fig.~\ref{fig: oddcombination}. 

Thus we neglected odd combination of phonons, this leads to $D = D_0/\sqrt{N}$, where factor of $1/\sqrt{N}$ is renormalization factor \cite{ZiyanLi2023}, resulting in $D = D_0/\sqrt{3}$ for 3 layers systems and $D = D_0/2$ for 4 layers systems. We take $\lambda = 3.25$ eV$ \text{\r{A}}^{-2}$, $\mu = 9.44$ eV$ \text{\r{A}}^{-2}$ and $D_0 = 20$ eV. We consider the static limit of phonon-mediated electron-electron interaction: $\mathcal{V}_{ph}(\vb{q},\omega) \approx \mathcal{V}_{ph}(\vb{q},\omega = 0) = \frac{D^2}{\lambda+2\mu}$.\\

The effective screened potential responsible for the superconducting pairing is given by,
\begin{equation}
    \mathcal{V}_{scr}(\vb{q}) =\mathcal{V}_{C}(\vb{q})\left(I - \mathcal{V}_{C}(\vb{q})\chi_{\text{ph}}(\vb{q})\right)^{-1}.
\end{equation}
For the analysis of the superconducting instability it is convenient to begin with the linearized gap equation,
\begin{equation}
    \Delta(\vb{k}) = -\frac{k_{B}T}{\Omega}\sum_{\vb{q},\omega,i',j'}{\mathcal{V}_{scr}(\vb{k}-\vb{q})\mathcal{G}^{i,i'}(\vb{q},\omega)\mathcal{G}^{j,j'}(-\vb{q},-\omega)\Delta(\vb{q})},
    \label{eq: linearizedgap}
\end{equation}
where $K_{B}$ is the Boltzmann constant, $T$ is the temperature, $\mathcal{G}$ is the electronic Green's function in the normal state and $\omega$ are fermionic Matsubara frequencies. Following an analog procedure as in Refs.~\cite{JimenoPozo2023, ZiyanLi2023} the linearized gap equation can be solved self-consistently as an eigenvalue problem for the kernel,
\begin{equation}
    \begin{split}
        \Gamma_{n_{1},n_{2},m_{1},m_{2}}^{\alpha,\alpha^{\prime}}(\vb{k},\vb{q}) = -\frac{1}{\Omega}&\sqrt{\frac{f(-\epsilon_{m_{2},\vb{k},\alpha}) - f(\epsilon_{m_{1},\vb{k},\alpha})}{\epsilon_{m_{1},\vb{k},\alpha} + \epsilon_{m_{2},\vb{k},\alpha}}}\sqrt{\frac{f(-\epsilon_{n_{2},\vb{q},\alpha}) - f(\epsilon_{n_{1},\vb{q},\alpha})}{\epsilon_{n_{1},\vb{q},\alpha} + \epsilon_{n_{2},\vb{q},\alpha}}}\times \\ & \langle\ \Psi_{n_{1},\vb{q},\alpha}||\Psi_{m_{1},\vb{k},\alpha}\rangle^\dagger\cdot \mathcal{V}_{scr}(\vb{k}-\vb{q}) \cdot \langle\Psi_{n_{2},\vb{q},\alpha}||\Psi_{m_{2},\vb{k},\alpha}\rangle .
    \end{split}
    \label{eq: GapEquation}
\end{equation}
We further note that the derivation of susceptibility and linearized gap equation using our methodology is equivalent to the result obtained by Fourier transforming the real-space Green function~\cite{cea21Coulomb}. Note that the above equation is written in terms of the overlap vector which contains information about the strength of the charge density. The onset of superconductivity is determined by the divergence of the Cooper pair correlation function which takes places when the maximum eigenvalue of the kernel $\Gamma_{n_{1},n_{2},m_{1},m_{2}}^{\alpha,\alpha^{\prime}}$ takes a value of $1$. To compute the kernel we consider those electronic states close to the Fermi energy within a range of $10$ meV, and then we change the temperature until the maximum eigenvalue reaches unity.

\section{Order Parameter, Fermi Surfaces and Matrix Elements of the Screened Coulomb Potential in TBG}
\begin{figure}[ht!]
    \centering
    \includegraphics[width =0.9 \textwidth]{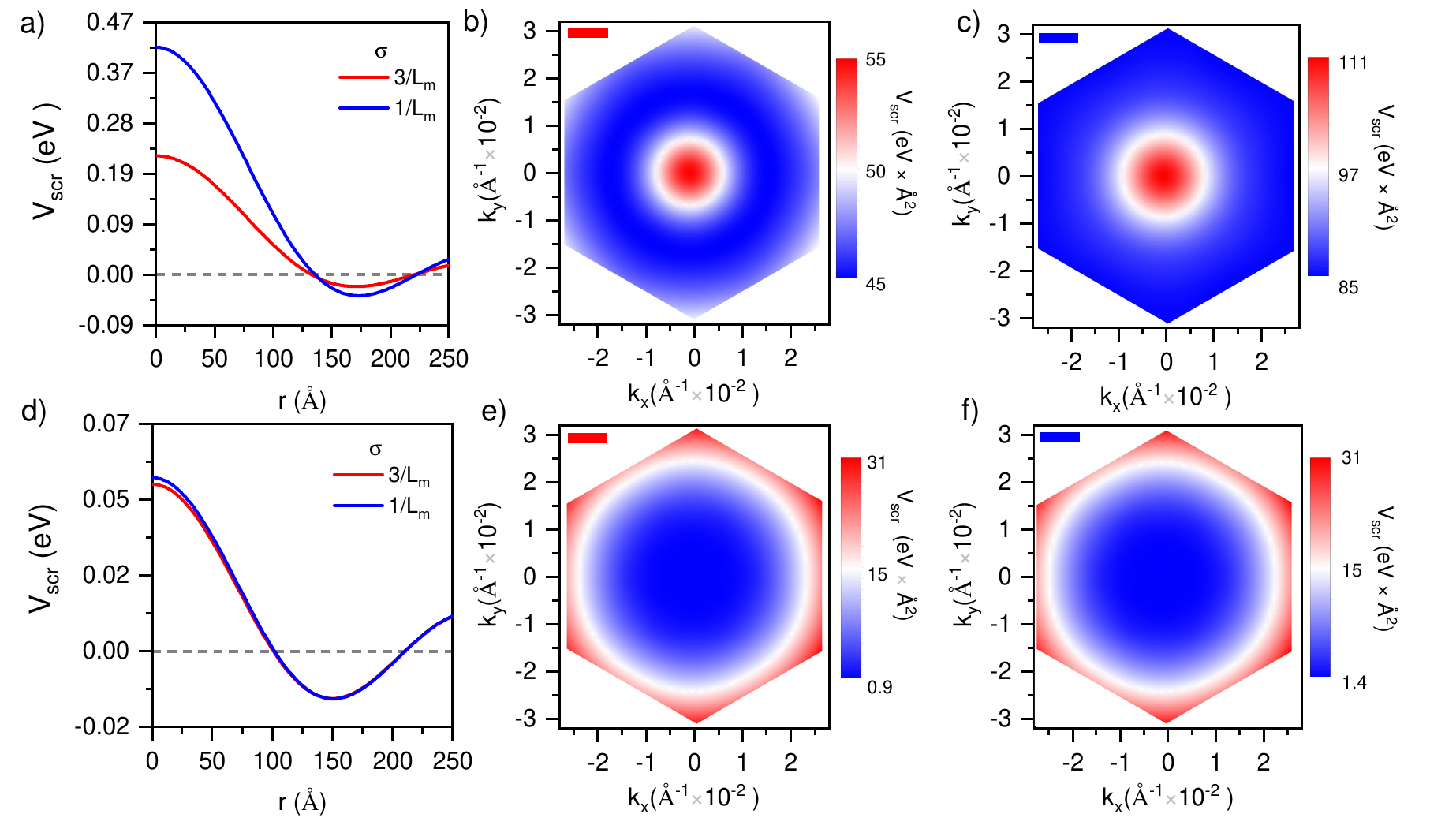}
    \caption{Screened Coulomb potential obtained with a Gaussian approximation for TBG at $\nu=-2.5$ and $V_H = -4.85$~meV. Top and bottom rows are for the Gaussian approximations in Eq.~\ref{eq: Gaussian1} and Eq.~\ref{eq: Gaussian2}, respectively. The Gaussian broadening is indicated in each panel in a) and d). The corresponding momentum space representations of screened potential are shown in the density plots.}  
    \label{fig: Real Space}
\end{figure}

\begin{figure}[ht!]
    \centering
    \includegraphics[width = \textwidth]{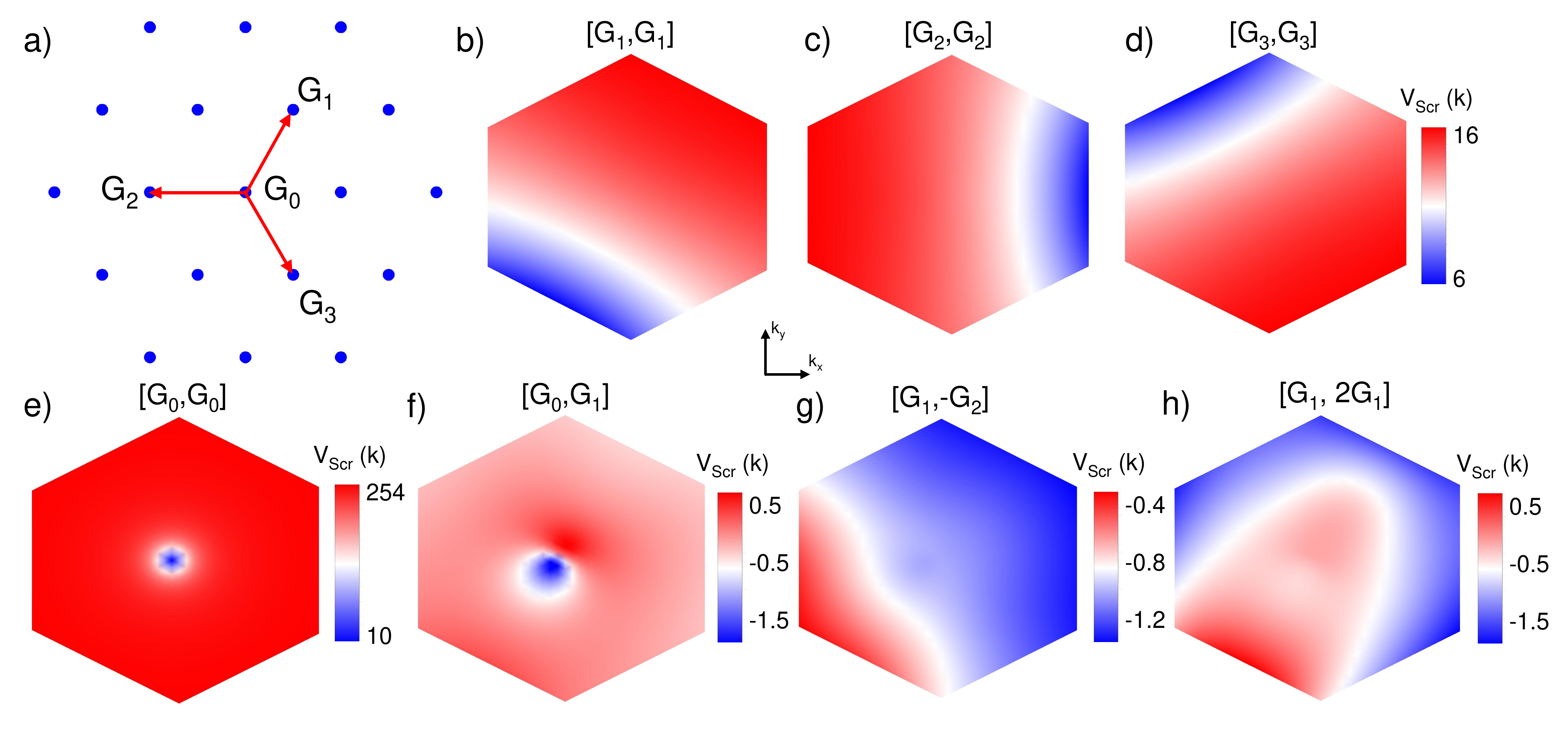}
    \caption{Matrix elements of the screened Coulomb potential $V(\vb{q})_{\vb{G_i},\vb{G_j}}$ with $\vb{q}=\{k_x,k_y \}$. In a) we show the $\vb{G}$ vectors in the reciprocal space. The $C_3$ symmetry of the Coulomb potential is shown for the cases of b) $V_{\vb{G_1},\vb{G_1}}(\vb{q})$, c) $V_{\hat{C_3}\vb{G_1},\hat{C_3}\vb{G_1}}(\vb{q})$ and d) $V_{\hat{C_3}^2\vb{G_1},\hat{C_3}^2\vb{G_1}}(\vb{q})$. Panel e) display the Coulomb potential for $V_{\vb{G_0},\vb{G_0}}(\vb{q})$. Panels f) to h) display the contribution for the states with $\vb{G_i} \neq \vb{G_j}$. In the top of each panel the matrix element is indicated. The units of $V_{scr}$ are meV.} 
    \label{fig: Vqplots}{
}\end{figure}

\begin{figure}[ht!]
\centering
\includegraphics[width =0.8 \textwidth]{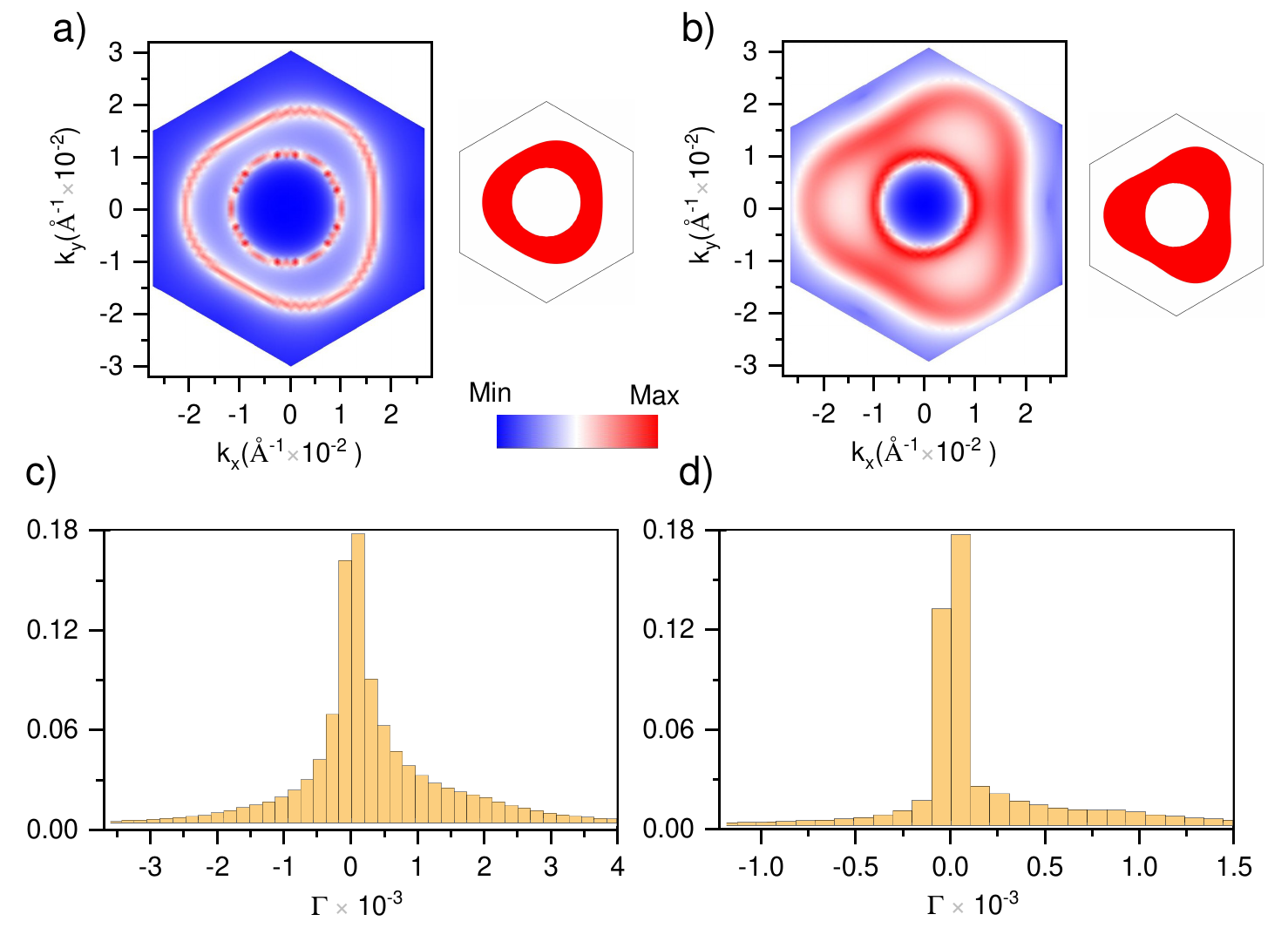}
\caption{Order parameter projected onto the valence band throughout the
BZ for TBG at $1.1^\circ$ and different Hartree potential strength. In a) we set $V_H = -4.85$~meV with $\nu=-2.5$ and in b) $V_H = -9.56$~meV with $\nu=-2.81$. In c) and d) we show the probability distribution of the real part of the matrix elements of the $\Gamma$ matrix in Eq.~\ref{eq: GapEquation}. The critical temperatures are $1250$ and $500$ mK for the potentials in a) and b), respectively. The corresponding Fermi surface is shown in each panel.}  
    \label{fig: Oparameter}
\end{figure}

In a moiré multilayer, the dielectric function, $\epsilon_{\vec{G},\vec{G}'} ( \vec{q} , \omega )$, and the screened interaction, ${\cal V}^{scr}_{\vec{G},\vec{G}'} ( \vec{q} , \omega )$ are matrices which depend on the reciprocal lattice vectors $\vec{G}, \vec{G}'$. As a result, and unlike the case of the untwisted systems studied in Ref.~\cite{JimenoPozo2023}, a straightforward real space Fourier transform of the screened potential cannot be calculated. A simplified average over reciprocal lattice vectors can be made by projecting the screened potential onto a localized function defined within the moiré unit cell, $\Psi ( \vec{r} ) = \sum_{\vec{G}} \alpha_{\vec{G}} e^{i \vec{G} \vec{r}}$. We have defined a Gaussian wavefunction, as in Ref.~\cite{Ishizuka_2021,Dong2023}, and calculated:
 \begin{equation}
{\cal V} ( \vec{q} ) = \sum_{ \vec{G}' , \vec{G}'''} \langle\alpha |M(\vec{G}')|\alpha\rangle \langle \alpha|M(\vec{G}''') |\alpha\rangle {\cal V}^{scr}_{G' , G'''} ( \vec{q} , \omega = 0 )
\label{eq: Gaussian Ref}
\end{equation}
where $|\alpha\rangle =(\alpha_{\vec{G_0}},\cdots,\alpha_{\vec{G}_N}) $, $M(\vec{G})$ is a connection matrix and 
\begin{equation}
     \alpha_{\vec{G}} = \frac{A}{\sqrt{2\pi}\sigma}e^{-\frac{|\vec{G}|^2}{2\sigma^2}},
     \label{eq: Gaussian1}
\end{equation}
 is the Gaussian wavefunction. An alternative to Eq.~\ref{eq: Gaussian1} is given by
 \begin{equation}
     \alpha_{\vec{G}} = \frac{A'}{\sqrt{2\pi}\sigma}|G| e^{-\frac{|\vec{G}|^2}{2\sigma^2}},
     \label{eq: Gaussian2}
\end{equation}
where $A$ is defined so that $1 = \sum_{\vec{G}} \alpha^*_{\vec{G}} \alpha_{\vec{G}}$. The real space Coulomb potential is then obtained by Fourier transforming Eq.~\ref{eq: Gaussian Ref}. Results are shown in Fig.~\ref{fig: Real Space} for TBG at $1.1^\circ$. We consider two values of the Hartree potential $V_H = -4.85$ meV with $\nu=-2.5$ and in $V_H = -9.56$ meV with $\nu=-2.81$.  
The order parameter and the corresponding Fermi surface are shown in Fig.~\ref{fig: Oparameter}. We note that the expression in Eq.~\ref{eq: Gaussian Ref} implicitly takes into account the Umklapp processes. To properly visualize the screened coulomb potential we also display the matrix elements in Fig.~\ref{fig: Vqplots}, where is evident that Umklapp processes makes some regions attractive. 

\section{Symmetry of Umklapp process}
\label{sec: Symmetry}
In this section, we present symmetry properties of the Umklapp process, we consider a $C_3$ rotational symmetry which is common for the systems studied in this work. In calculating susceptibility, we need to evaluate the overlap of an state between momentum $\vb{k}$ and momentum $\vb{k}+\vb{q}$. If $\vb{k}+\vb{q}$ is out of the mBZ, we can connect the state $\vb{k} + \vb{q}$ with an state $\vb{k}+\vb{q} - \vb{G}$, where $\vb{k}+\vb{q} - \vb{G}$ is within the mBZ and $\vb{G}$ is a reciprocal lattice vector. The relation between the state at $\vb{k}+\vb{q} - \vb{G}$ and the state at $\vb{k}+\vb{q}$ can be determined by considering the symmetry of the Hamiltonian.
\begin{figure}[ht!]
    \centering
    \includegraphics[scale = 0.5]{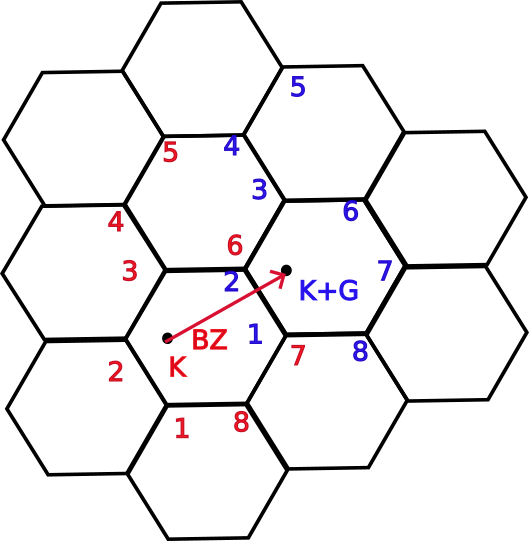}
    \caption{Periodic boundary conditions in the Brillouin Zone, the index of Dirac cones for $H(\vb{k})$ is labeled in red and that for $H(\vb{k}+ \vb{G})$ is labeled in blue.} 
    \label{fig: translational symmetry}
\end{figure}
For an state at $\vb{k}$, we can construct a Hamiltonian $H(\vb{k})$ and  $H(\vb{k}+\vb{G})$ in a similar way. However, when constructing the matrix $H(\vb{k}+\vb{G})$, we can change the indexes of the Dirac cones such that the matrix form of $H(\vb{k})$ and $H(\vb{k}+\vb{G})$, which are written in their own basis separately, is the same. Then we found that the eigenvalues at both momentum are the same, except that the eigenstates are defined in a different basis. To change the basis, we can use the connection matrix to do a transformation and we get:
\begin{align}
    \epsilon_{\vb{k}+\vb{G}}& =  \epsilon_{\vb{k}} \\
    |\psi_{\vb{k}+\vb{G}}\rangle &= M(\vb{G})|\psi_{\vb{k}}\rangle.
    \label{SIeq: statemoveG}
\end{align}
For the screened potential, which is a matrix, we need to do a similarity transformation
\begin{equation}
    V(\vb{k}+\vb{G}) = M(\vb{G})V(k)M(\vb{G})^T.
\end{equation}
The form factor is an important quantity for describing correlation effects in moir\'e systems.
\begin{equation}
\begin{split}
     \lambda_{m,n,\alpha}(k+q,k,G) = \sum_{\beta,\vb{G'}} u^*_{\alpha,\beta,m,\vb{G'+G}}(\vb{k+q})u_{\alpha,\beta,n,\vb{G'}}(\vb{k}) =  \langle \Psi_{m,\mathbf{k+q}}|M(\vb{G})|\Psi_{n,\mathbf{k}}\rangle =  \langle \Psi_{m,\mathbf{k+q+G}}|\Psi_{n,\mathbf{k}}\rangle. 
\end{split}
\end{equation}
We see that the form factor is nothing else but the inner product of the states, but the non-trivial terms appear when we take the two states belonging to different Voronoi cells. 
Now we show that the form factor doesn't change under the translation of arbitrary reciprocal vector in the language of the connection matrix, this allows us to evaluate the form factor in any primitive cell of momentum space.
\begin{equation}
\begin{split}
     \lambda_{m,n,\alpha}(k+q+G',k+G',G) = &\langle \Psi_{m,\mathbf{k+q+G'}}|M(\vb{G})|\Psi_{n,\mathbf{k+G'}}\rangle \\
      =  &\langle \Psi_{m,\mathbf{k+q}}|M^\dagger(\vb{G'})M(\vb{G})M(\vb{G'})|\Psi_{n,\mathbf{k}}\rangle \\
      = & \langle \Psi_{m,\mathbf{k+q}}|M(\vb{-G'})M(\vb{G})M(\vb{G'})|\Psi_{n,\mathbf{k}}\rangle \\
      = & \langle \Psi_{m,\mathbf{k+q}}|M(\vb{-G'} + \vb{G} + \vb{G'})|\Psi_{n,\mathbf{k}}\rangle \\
      =  & \langle \Psi_{m,\mathbf{k+q}}|M(\vb{G})|\Psi_{n,\mathbf{k}}\rangle \\
      = & \lambda_{m,n,\alpha}(k+q,k,G) \\
\end{split}
\end{equation}
Where the first equation is the definition. We use the transformation rule of the states vector in the second equation. In the third equation we use the relation $M^\dagger(\vb{G}) = M(-\vb{G})$, this can be easily checked from the definition of the connection matrix. In the fourth  equations, we apply the abelian nature of the connection matrix since  $M(\vb{G})M(\vb{G'}) = M(\vb{G}+\vb{G'}) $. This can also be checked from the definition. We see that the connection matrix offers great convenience in studying correlation effects in moir\'e systems. We write again the susceptibility, 
\begin{equation}
\begin{split}
\left[\chi_{0}(\boldsymbol{q})\right]_{\boldsymbol{G}_{i,}\boldsymbol{G}_{j}}&=\frac{4}{\varOmega}\sum_{\boldsymbol{k},m,n,\alpha}\frac{f(\epsilon_{n,\boldsymbol{k}})-f(\epsilon_{m,\boldsymbol{k}+\boldsymbol{q}})}{\epsilon_{n,\boldsymbol{k}}-\epsilon_{m,\boldsymbol{k}+\boldsymbol{q}})}
\left\langle \Psi_{n,\boldsymbol{k+G_i}}|\Psi_{m,\boldsymbol{k+q}}\right\rangle \left\langle \Psi_{m,\boldsymbol{k+q}}\right|\Psi_{n,\boldsymbol{k+G_j}}\rangle,  \\
\end{split}
\end{equation}
where $|\Psi_{n,\boldsymbol{k+G}}\rangle$ is the eigenstate of a band $n$ at momentum $\vb{q+G}$. From the spirit of the connection matrix, we also have a $\hat{C}_3$ transformation matrix, defined as:
\begin{equation}
    M(\hat{C}_3)_{ij} =    \begin{cases}
       \text{$I_{\beta}$} &\text{ if $\hat{C}_3\vb{G}_i = \vb{G}_j$},\\
       \text{$0_{\beta}$} &\text{ if $\hat{C}_3\vb{G}_i \neq \vb{G}_j$}
    \end{cases}
\end{equation}
The $C_3$ symmetry of screened Coulomb potential is also expressed as, 
\begin{equation}
    V(\hat{C}_3\vb{k}) = M(\hat{C}_3)V(k)M(\hat{C}_3)^T \iff  V(\vb{k})_{\vb{G}_1,\vb{G}_2} = V(\hat{C}_3\vb{k})_{\hat{C}_3\vb{G}_1,\hat{C}_3\vb{G}_2}  
\end{equation}
where in the latter expression we explicitly write the matrix elements. The numerical result is shown in Fig.~\ref{fig: Vqplots}b)-d) for TBG.

\section{Contributions to Superconductivity}
\label{sec:Contributions}
\begin{figure}
     \centering
     \includegraphics[scale=0.40]{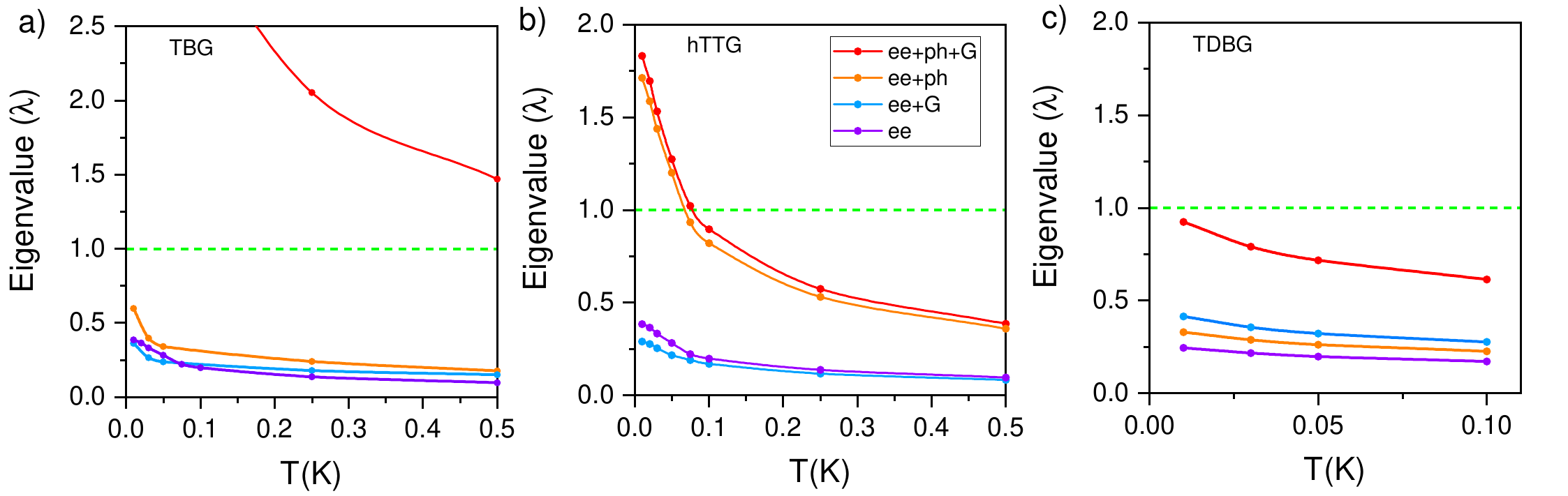}
     \caption{Evolution of the maximum eigenvalue, $\lambda(T)$ in Eq.~\ref{eq: GapEquation}, as a function of temperature for a) TBG, b) hTTG and c) TDBG. The intersection with the green line gives the critical temperature.
     }
     \label{fig:Fitting}
 \end{figure}

We now consider the relative strength of different contributions to the superconducting instability in TBG, TDBG and hTTG. We compute the leading eigenvalue of the kernel Eq.~\ref{eq: GapEquation} as a function of the temperature for the different combinations of direct electron-electron interactions (ee), phonon mediated electron-electron interactions (ph) and Umklapp processes (G). The results, shown in Fig.~\ref{fig:Fitting}, indicate that there is a rich interplay between these contributions to drive superconductivity, although in general it is clear that electron-electron interactions alone can not lead to a reasonable critical temperature in twisted graphene stacks. In the case of hTTG at its magic-angle, Fig.~\ref{fig:Fitting}b), we realize that the Umklapp contribution does not impact the base effect of electron-electron interaction, while the electron-phonon dressing is crucial to establish a critical temperature of the order of $T_{c}\approx 60$ mK. The posterior inclusion of Umklapp processes finally sets $T_{c}\approx 80$ mK. For TDBG, Fig.~\ref{fig:Fitting}c), both the phonon dressing or the Umklapp processes alone do not change meaningfully the underlying screening due to electron-electron interactions. Only once that both contributions are considered on top of the purely electronic interactions the critical temperature grows up to $T_{c}\approx 10$ mK. The tendency in TBG, Fig.~\ref{fig:Fitting}a), is in stark contrast to hTTG while resembling the TDBG situation but with a huge impact of the joined effect of electron-phonon interaction and Umklapp processes. The strong impact of Umklapp processes in the superconducting phase of TBG is reflected by the severe band distortions induced by the Hartree potential.

\section{Superconductivity with broken-symmetry parent state}
\label{sec:sc_brokensym}
\begin{figure}
    \centering
    \includegraphics[scale=0.60]{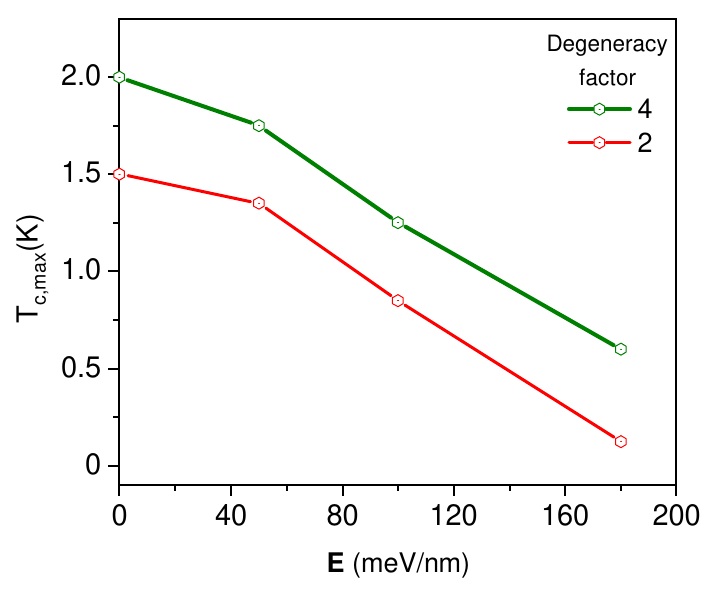}
    \caption{Maximum critical temperature of TBG as a function of a perpendicular electric field for different flavour degeneracy values in Eq.~\ref{eq:SusceptibilityMoire}. 
    }
    \label{fig:brokenSym}
\end{figure}

We now consider the effect of a parent state with a broken flavour symmetry. We approximate this effect by using a degeneracy factor of 2 instead of 4 in the electronic susceptibility in Eq.~\textcolor{red}{1}, c.f. Refs.~\cite{JimenoPozo2023,ZiyanLi2023}. In Fig.~\ref{fig:brokenSym}, we show the maximum critical temperature of TBG as a function of the external electric field for the cases when the degeneracy is 4 or 2. We find a reduction of the superconducting phase due to the symmetry breaking, ultimately leading to a decrease of the critical temperature. Surprisingly, this effect is opposite to our findings for non-twisted systems, in which the reduction of symmetry leads to a significant enhancement of the critical temperature~\cite{JimenoPozo2023, ZiyanLi2023}. However, this behavior is in agreement with recent experimental observations~\cite{Dutta2024}. In the case of magic-angle hTTG, we find a moderate reduction in the critical temperature, from $80$ mK to $50$ mK, when the degeneracy factor is changed. In TDBG, a qualitatively similar reduction of the eigenvalue is observed, suggesting a notable impact on the critical temperature. The non-trivial dependency of the superconducting phase on the appearance of broken-symmetry states seems to be quite different between twisted and non-twisted graphene stacks, and given the considerable aid of WSe$_{2}$ in stabilizing superconductivity in these systems, we consider this to be an issue that requires further investigation.

\section{Adjustable and fixed parameters in the KL-RPA framework}
\label{sec:fixed_parameters}

In this section, we discuss both the tunable and non-tunable parameters of the KL-RPA framework employed in our calculations. The KL-RPA framework begins with the computation of the electronic susceptibility, Eq.~\textcolor{red}{1}, that depends on the chemical potential and the details of the electronic structure. These are determined with the continuum models described in Sec.~\ref{sec:ContinuumTDBG} and Sec.~\ref{sec:ContinuumTTG}. The continuum models depend on a set of well-known fixed parameters summarized in Tab.~\textcolor{red}{I} and other variable parameters such as the external electrostatic potential $\Delta V$ and the scaling coupling parameter $\kappa_{h}$. In the calculation of the electronic susceptibility these variable parameters are fixed to contrast their impact in the superconducting instability, thus the only actual tunable parameters are the chemical potential $\mu$ and the temperature $T$. \\

The next step in the KL-RPA framework is to compute the electronic susceptibility screened by the electron-phonon potential following Eq.~\textcolor{red}{2}. The electron-phonon potential is given in Eq.~\ref{eq:electron_phonon_pot} and its static limit $\omega\to 0$ is considered in the calculations, being totally defined by the Lamé coefficients and the deformation potential. These parameters are fixed to the values summarized in Tab.~\textcolor{red}{I}, therefore no tunable parameters appear in this step. The screened electronic susceptibility and the Coulomb potential determine the final screened potential in Eq.~\textcolor{red}{3}. This potential is being used in the calculation of the superconducting kernel. The Coulomb potential, given in Eq.~\ref{eq:bare_coulomb_pot}, depends on the distance between the system and the metallic gates and on the relative dielectric constant which are fixed during the calculations to a value summarized in Tab.~\textcolor{red}{I}. At this point the only tunable parameters keep being $\mu$ and $T$.\\

The last step is to construct the superconducting kernel, Eq.~\ref{eq: GapEquation}, that depends on the screened potential and the details of the electronic structure. The onset of superconductivity is defined by the conditions that make one the leading eigenvalue of the kernel, realizing a solution for the linearized gap in Eq.~\ref{eq: linearizedgap}. In the KL-RPA framework the superconducting conditions are fully determined by the set of fixed of parameters summarized Tab.~\textcolor{red}{I} and by the chemical potential, $\mu$, and temperature $T$.\\


\begin{thebibliography}{113}
\bibitem [{gol()}]{goldstone}%
  \BibitemOpen
  \href@noop {} {}\bibinfo {note} {Note that we do not include electron-hole
  ladder diagrams which can describe soft modes due to the vicinity to a
  polarized phase, see, for instance~\cite{DCL22}.}\BibitemShut {Stop}%
\end{thebibliography}
\end{document}